\documentclass[showpacs,reprint,a4paper]{revtex4-1}

\usepackage{amsmath}
\usepackage{amssymb}
\usepackage{graphicx}
\usepackage{tikz}
\usepackage{slashed}
\usepackage{multirow}
\usepackage{hhline}
\usepackage{subcaption}
\usepackage{caption}
\usepackage{float}
\usepackage{feynmp}

\begin{document}

\title{Chiral pair of Fermi arcs, anomaly cancelation, and spin or valley Hall effects in inversion symmetry-broken Weyl metals}
\author{Iksu Jang and Ki-Seok Kim}
\affiliation{Department of Physics, POSTECH, Pohang, Gyeongbuk 790-784, Korea}
\date{\today}

\begin{abstract}
Anomaly cancelation has been shown to occur in time-reversal symmetry-broken Weyl metals, which explains the existence of a Fermi arc. We extend this result in the case of inversion symmetry-broken Weyl metals. Constructing a minimal model that takes a double pair of Weyl points, we demonstrate the anomaly cancelation explicitly. This demonstration explains why a chiral pair of Fermi arcs appear in inversion symmetry-broken Weyl metals. In particular, we find that this pair of Fermi arcs gives rise to either ``quantized" spin Hall or valley Hall effects, which corresponds to the ``quantized" version of the charge Hall effect in time-reversal symmetry-broken Weyl metals.
\end{abstract}

\maketitle

\section{Introduction}

Anomaly cancelation is the mechanism to explain the existence of a gapless surface state, topologically protected \cite{Callan-Harvey}. For example, the existence of a chiral edge mode in the integer quantum Hall effect is understood as follows \cite{Quantum_Hall_Review_Stern}. The chiral edge state suffers gauge anomaly, which means that the U(1) current is not conserved. On the other hand, the Chern-Simons term is not invariant under the gauge transformation in the presence of a boundary. It turns out that the gauge anomaly at the boundary is canceled exactly by the gauge non-invariant term of the Chern-Simons theory in the bulk. As a result, a topological term with a gapless boundary mode consists of a topological field theory consistently.

A Weyl metal state may be regarded as a three dimensional generalization of an integer quantum Hall phase \cite{WM1,WM2,WM3,WM4,WM_Review1,WM_Review2,WM_Review3,TSB_WM1,TSB_WM2}. The Chern-Simons term is replaced with a topological-in-origin $\bm{E} \cdot \bm{B}$ term. The ``axion" $\theta$ field corresponding to the Hall conductance in the integer quantum Hall effect is proportional to the displacement from a reference point and its gradient is nothing but an applied magnetic field to describe the momentum-space distance between a pair of Weyl points in the case of time-reversal symmetry-breaking. A Fermi arc state corresponds to the chiral edge mode, responsible for the existence of an anomalous Hall effect. As the gauge anomaly from the edge state must be canceled by the gauge non-invariant term from the Chern-Simons term in the integer quantum Hall state, the gauge anomaly from the Fermi arc is also canceled by a gauge non-invariant contribution at the boundary from the inhomogeneous axion term. As a result, the topological-in-origin inhomogeneous $\theta$ term with the Fermi arc state gives a consistent ``topological" field theory for the time-reversal symmetry-broken Weyl metal phase, where contributions from massless Weyl-fermion excitations should be taken into account, of course.

In this study we extend the anomaly cancelation of a time-reversal symmetry-broken Weyl metal state into that of an inversion symmetry-broken Weyl metal phase. The minimal model of the time-reversal symmetry-broken Weyl metal state is given by a pair of Weyl points, where the momentum-space distance between the pair of Weyl points is the gradient $\theta$ proportional to the applied magnetic field. On the other hand, that of the inversion symmetry-broken Weyl metal phase is given by a double pair of Weyl points, where the momentum-space distance between each pair of Weyl points is determined by the strength of the inversion symmetry breaking. Based on this minimal model, we demonstrate the anomaly cancelation explicitly. This demonstration explains why a ``chiral" pair of Fermi arcs instead of a Fermi arc with definite chirality appear in inversion symmetry-broken Weyl metals.

One may point out that the explicit demonstration for the anomaly cancelation in the inversion symmetry-broken Weyl metal phase does not give any novel conceptual aspect, compared with that in the time-reversal symmetry-broken Weyl metal state. However, we claim that there are no concrete calculations to show the anomaly cancelation in the inversion symmetry-broken Weyl metal state. In addition, we emphasize that there exists novel physics in the anomaly cancelation of the inversion symmetry-broken Weyl metal phase. Since time reversal symmetry is preserved, a ``quantized" version of the anomalous Hall effect resulting from the Fermi arc cannot appear. Instead, we find that this pair of Fermi arcs give rise to either ``quantized" spin Hall or valley Hall effects, which may be regarded to be a ``generalized" version of the two dimensional quantum spin or valley Hall effect. In this respect we believe that our explicit demonstration serves as a meaningful reference in understanding the ``chiral" pair of Fermi arc states in various inversion symmetry-breaking Weyl metals \cite{ISB_WM1,ISB_WM2,ISB_WM3,ISB_WM4,ISB_WM5,ISB_WM6,ISB_WM7}.

\section{A review on the anomaly cancelation in the time-reversal symmetry-broken Weyl metal state}

\subsection{An effective minimal model for time-reversal symmetry-broken Weyl metals}

A minimal model for time-reversal symmetry-broken Weyl metals is given by \cite{Disordered_Weyl_Metal1,Disordered_Weyl_Metal2}
\begin{align}
S_{WM}&=\int d^4 x \bar{\Psi}(x) (\gamma_0\partial_0+i\gamma^k\partial_k-\mu\gamma^0-c_\mu \gamma^\mu\gamma^5) \Psi(x) ,
\end{align}
where $\gamma^0=\Big(\begin{array}{cc} 0 & 1 \\ 1 & 0 \end{array}\Big)$, $\gamma^k=\Big(\begin{array}{cc} 0 & \sigma^k \\ -\sigma^k & 0 \end{array} \Big)$, and $\gamma^5=i\gamma^0\gamma^1\gamma^2\gamma^3$. $c^\mu=(c^0,\mathbf{c})$ is the chiral gauge field, where $c^0$ is the chiral chemical potential and $\mathbf{c}$ is the momentum-space distance between a pair of Weyl points. $\mu$ is the chemical potential. Here, we focus on $\mu = 0$ and $c^0 = 0$.

Introducing $\gamma^4=-i\gamma^0$ into the above action, we have a simplified form
\begin{align}
S_{WM}=\int d^4 x \bar{\Psi}(x) i\gamma^\mu( \partial_\mu +iA_\mu+ic_\mu\gamma^5) \Psi(x) . \label{Weyl_Metal_TRSB}
\end{align}
Here, we have $\mu=1,2,3,4$. $\gamma^\mu$ is anti-Hermitian, satisfying $\{\gamma^\mu,\gamma^\nu\}=-2\delta^{\mu\nu}$. We take into account the U(1) gauge field $A_\mu$.

\subsection{An axion term}

The Weyl-metal action Eq. (\ref{Weyl_Metal_TRSB}) suffers chiral anomaly \cite{Nielsen_Ninomiya}, given by
\begin{align}
\partial_{\mu} \bar{\Psi}(x) \gamma^\mu \gamma^5 \Psi(x) = \frac{1}{16\pi^2} \epsilon^{\mu\nu\alpha\beta} F_{\mu\nu} F_{\alpha\beta} .
\end{align}
Although it is straightforward to derive this anomaly equation based on the Fujikawa's method \cite{Anomalies in QFT,Burkov}, we show our derivation explicitly in Appendix \ref{Axionic-action-TRB WM} in order to clarify the way of regularization. The resulting axionic action is
\begin{gather}
S_{ax}= - \frac{i}{16\pi^2} \int d^4x (c_\mu x^\mu) \epsilon^{\mu\nu\alpha\beta} F_{\mu\nu} F_{\alpha\beta} . \label{AxionicAction}
\end{gather}

\subsection{Surface states}

Following Goswami and Tewari \cite{TRB WM Anomaly}, we obtain a gapless surface state, referred to as a Fermi arc state. Detailed calculations are shown in Appendix \ref{Appendix:surface-state-TRB Wm}. The resulting localized wave function at one surface of the Weyl metal phase is given by
\begin{align}
 \psi_{k_y,k_z}(x,y,z)&=Ae^{ik_y y+ik_z z}\left(\begin{array}{c}1 \\ i \\ -\frac{m}{\sqrt{k_z^2+m^2}-k_z}\\ i\frac{m}{\sqrt{k_z^2+m^2}-k_z}\end{array}\right)\nonumber \\
 &\times e^{(-c\theta(x)+\sqrt{k_z^2+m^2})x}.
\end{align}
where $A = \Big(\frac{(\sqrt{k_z^2+m^2}-k_z)^2\sqrt{k_z^2+m^2}(c-\sqrt{k_z^2+m^2})}{2c(k_z^2+m^2-k_z\sqrt{k_z^2+m^2})}\Big)^{1/2}$ is a normalization constant and $E_{k_y}=k_y$ is an eigen value of this surface state. $y$ and $z$ define the surface coordinate and $x$ describe the coordinate along the bulk direction. The chiral gauge field is given along the $z$ direction. For a state localized near the surface to exist, $k_z$ should satisfy the following condition of $-\sqrt{c^2-m^2}<k_z<\sqrt{c^2-m^2}$. If $m^2>c^2$ is fulfilled, there are no surface states. It is important to realize that this surface state has definite chirality, given by $\bar{\gamma} \psi_{k_y,k_z} = - \psi_{k_y,k_z}$ with the chirality operator $\bar{\gamma} = \gamma^0 \gamma^2$ (Appendix \ref{Appendix:surface-state-TRB Wm}3).

\subsection{An effective Hamiltonian for the Fermi arc}

Let us now establish an effective Hamiltonian for the Fermi arc state. We introduce a surface project operator as follows
\begin{gather}
%
%
P_{edge} \equiv \sum_{k_y}\sum_{-\sqrt{c^2-m^2}<k_z<\sqrt{c^2-m^2}}|\psi_{k_y,k_z}\rangle\langle \psi_{k_y,k_z}| ,
\end{gather}
where we have $\langle x,y,z|\psi_{k_y,k_z}\rangle = \psi_{k_y,k_z}(x,y,z)$. Then, we construct an effective surface Hamiltonian in the following way
\begin{widetext}
\begin{align}
H_{eff}&=P_{edge}HP_{edge}=\sum_{k_y}\sum_{\tilde{k}_z}|\psi_{k_y,k_z}\rangle k_y \langle \psi_{k_y,k_z}|\nonumber \\
&=\int dx'\int dy'\int dx\int dy \sum_{k_y} \sum_{\tilde{k}_z} |x,y\rangle \langle x,y|\psi_{k_y,k_z}\rangle k_y \langle \psi_{k_y,k_z}|x',y'\rangle \langle  x',y'|\nonumber \\
%
%
&\approx \int dx'\int dy'\int dx\int dy\sum_{k_y} \sum_{\tilde{k}_z} |x,y,k_z\rangle(-i)\partial_{y}e^{ik_y(y-y')}\delta(x)\delta(x')\langle x',y', k_z|\nonumber\\
%
%
&=\int dy \sum_{\tilde{k}_z} |x=0,y,k_z\rangle (-i)\partial_{y}\langle x=0,y,k_z| ,
\end{align}
\end{widetext}
where $\tilde{k}_z$ means $k_z$ to satisfy $-\sqrt{c^2-m^2}<k_z<\sqrt{c^2-m^2}$. For simplicity, we assumed that the surface wave function is localized perfectly at the surface, i.e., $\langle x,y|\psi_{k_y,k_z}\rangle \sim \delta(x)$. This expression can be translated into
\begin{gather}
H_{eff}=\sum_{-\sqrt{c^2-m^2}<k_z<\sqrt{c^2-m^2}}\int dy \psi^\dagger_{k_z}(y)(-i)\partial_y \psi_{k_z}(y)
\end{gather}
in the second quantization language.

\subsection{Gauge anomaly in the Fermi arc state}

$(1+1)$ dimensional Dirac theory is given by
\begin{gather}
S=\int d^2x \bar{\Psi}(x)(\gamma_0\partial_0+i\gamma^1\partial_1)\Psi(x)
\end{gather}
in the Euclidean signature. Here, we have $\gamma^0=\sigma^1$ and $\gamma^1=i\sigma^2$. If we set $\gamma^2=-i\gamma^0$, we obtain
\begin{gather}
S=\int d^2 x \bar{\Psi}(x)i\gamma^\mu \partial_\mu\Psi(x)
\end{gather}
with $\mu=1,2$. Here, we have $g^{\mu\nu}=-\delta^{\mu\nu}$.

Let us gauge the above theory in the chiral gauge. Then, we obtain
\begin{gather}
S=\int d^2x \bar{\Psi}(x)i\gamma^\mu(\partial_\mu+ieA_\mu \mathcal{P}_-)\Psi(x) ,
\end{gather}
where $\mathcal{P}_-=\frac{1}{2}(1-\bar{\gamma})$ is a projection operator to select the chirality and $\bar{\gamma}=\gamma^0\gamma^1=i\gamma^2\gamma^1=-\sigma^3$ is the chirality matrix. Notice that we couple the U(1) gauge field only to the negative chirality sector. Recall that the edge mode in the above section has the negative chirality, i.e., $\bar{\gamma}\phi_{k_y,k_z}=-\phi_{k_y,k_z}$.
%
%
This effective action is invariant under the particular or ``partial" gauge transformation:
\begin{gather}
A_\mu(x)\rightarrow A_\mu+\partial_\mu\theta \nonumber \\
 \Psi\rightarrow e^{i\theta \mathcal{P}_-}\Psi(x),\;\; \bar{\Psi}(x)\rightarrow \bar{\Psi}(x)e^{-i\theta\mathcal{P}_+}.
\end{gather}

The U(1) gauge current, which is a Noether current resulting from the above partial gauge symmetry, is given by
\begin{gather}
j^\mu=\bar{\Psi}(x)\gamma^\mu P_-\Psi(x) .
\end{gather}
Classically, i.e., in the action level this gauge current is conserved. However, it turns out that this conservation law breaks down in the partition function level because of a quantum correction, referred to as gauge anomaly. This anomaly can be understood perturbatively in the one-loop quantum correction for the gauge-field propagator. See Appendix \ref{Appendix:1-loop-correction-TRB WM} for more details. The result is well known \cite{Anomalies in QFT}, given by
\begin{gather}
\partial_\mu j^{\mu}(x)=\frac{i}{4\pi}\epsilon^{\mu\nu}\partial_\mu A_\nu(x)=\frac{i}{8\pi}\epsilon^{\mu\nu}F_{\mu\nu} .
\end{gather}

One can express the gauge anomaly in terms of an effective action of the U(1) gauge field as follows
\begin{gather}
\frac{\delta W[A]}{\delta A_\mu}=-\langle \bar{\Psi}(x)\gamma^\mu \mathcal{P}_- \Psi(x) \rangle=-\langle j^{\mu}\rangle ,
\end{gather}
where the generating function is defined by $Z=\int \mathcal{D}\bar{\Psi}\mathcal{D}\Psi e^{-S[\bar{\Psi},\Psi,A]}\equiv e^{-W[A]}$.
Under the gauge transformation $A_\mu\rightarrow A_\mu \partial_\mu \eta(x)$, the generating function changes in the following way
\begin{align}
&\delta_\eta W[\mathcal{A}]\equiv W[\mathcal{A}+d\eta]-W[\mathcal{A}]\nonumber \\
&=\int d^2x \partial_\mu \eta(x)\frac{\delta W[A]}{\delta A_\mu} \nonumber \\
&=-\int d^2x \partial_\mu \eta(x)\langle j^{\mu}\rangle \nonumber \\
%
%
&=\int d^2x \eta(x)\frac{i}{8\pi}\epsilon^{\mu\nu}F_{\mu\nu} ,
\end{align}
where the gauge anomaly equation has been used.

Since $W[\mathcal{A}]$ is an effective action of only one $k_z$ sector, we should include all $k_z$ sectors in order to get the effective surface action of the Weyl metal phase
\begin{gather}
W_{WM}^{edge}[\mathcal{A}]=\sum_{-c<k_z<c}W[\mathcal{A}]=2cW[\mathcal{A}] .
\end{gather}
Here, we set $m=0$ for simplicity. As a result, we find the gauge anomaly of the Fermi arc state in the time-reversal symmetry-broken Weyl metal phase
\begin{align}
\therefore \delta_{\eta}W_{WM}^{edge}[\mathcal{A}]&=\frac{ic}{4\pi}\int dt dz \eta(x)\epsilon^{\mu\nu}F_{\mu\nu}\nonumber \\
&=\frac{ic}{2\pi}\int dt dz \eta(x)F_{zt} . \label{surfaceAnomaly}
\end{align}
Here, we did not take into account the role of disorder scattering for this gauge anomaly contribution. It would be quite an interesting study to investigate the role of disorder scattering for the Fermi arc state.

\subsection{Anomaly cancelation: Callan-Harvey mechanism}

Breakdown of the gauge invariance in the effective chiral surface state can be cured by anomaly inflow from the bulk effective action of the Weyl metal phase. This mechanism of anomaly cancelation is known as Callan-Harvey mechanism \cite{Callan-Harvey}. The Callan-Harvey mechanism has been already discussed in the time-reversal symmetry-broken Weyl metal phase \cite{TRB WM Anomaly}. However, we found a subtle issue for the derivation of the anomaly cancelation. Here, we provide a rigorous derivation for the anomaly cancelation based on the original paper \cite{Callan-Harvey}.

First, let us point out the subtle problem. One may start from an effective axionic action Eq. (\ref{AxionicAction}) with setting the chiral gauge field as $\mathbf{c}=c\Theta(x_1)\hat{z}$. Here, $\Theta(x_1)$ is the step function. The axion term is
\begin{align}
S_{ax}[A]&=\frac{i}{16\pi^2}\int d^4x \mathbf{c}\cdot\mathbf{x}\epsilon^{\mu\nu\alpha\beta}F_{\mu\nu}F_{\alpha\beta}\nonumber \\
&=\frac{i}{16\pi^2}\int d^4x cx_3\Theta(x_1)\epsilon^{\mu\nu\alpha\beta}F_{\mu\nu}F_{\alpha\beta}\nonumber \\
&=-\frac{i}{8\pi^2}\int d^4xc\epsilon^{\mu\nu\alpha\beta}[x_3 A_{\nu}F_{\alpha\beta}\delta(x_1)\delta_{1\mu}\nonumber \\
&+\Theta(x_1)\delta_{3\mu}A_\nu F_{\alpha\beta}] ,
\end{align}
where $\partial_x\Theta(x)=\delta(x)$ has been used. Under the gauge transformation $A_\mu\rightarrow A_\mu+\partial_\mu \eta(x)$, the variation of the effective action ($\delta_\eta S_{ax}\equiv S_{ax}[A_\mu+\partial_\mu\eta]-S_{ax}[A_\mu]$) is given by
\begin{align}
\delta_{\eta}S_{ax}&=-\frac{ic}{8\pi^2}\int d^4x\Big[\epsilon^{1\nu\alpha\beta}x_3F_{\alpha\beta}\delta(x_1) \nonumber \\
&+\epsilon^{3\nu\alpha\beta}\Theta(x_1)F_{\alpha\beta}\Big]\partial_\nu\eta\nonumber \\
&=\frac{ic}{8\pi^2}\int d^4x\Big[\epsilon^{1\nu\alpha\beta}F_{\alpha\beta}\Big(\delta_{\nu 3 }\delta(x_1)+x_3\partial_1\delta(x_1)\delta_{\nu 1}\Big)\nonumber \\
&+\epsilon^{3\nu\alpha\beta}\delta_{\nu 1}F_{\alpha\beta}\delta(x_1)\Big]\eta\nonumber \\
&=0 .
\end{align}
There does not exist the anomaly inflow to cancel the gauge anomaly of the Fermi arc state in this derivation.

\begin{figure}
\includegraphics[width=0.5\textwidth]{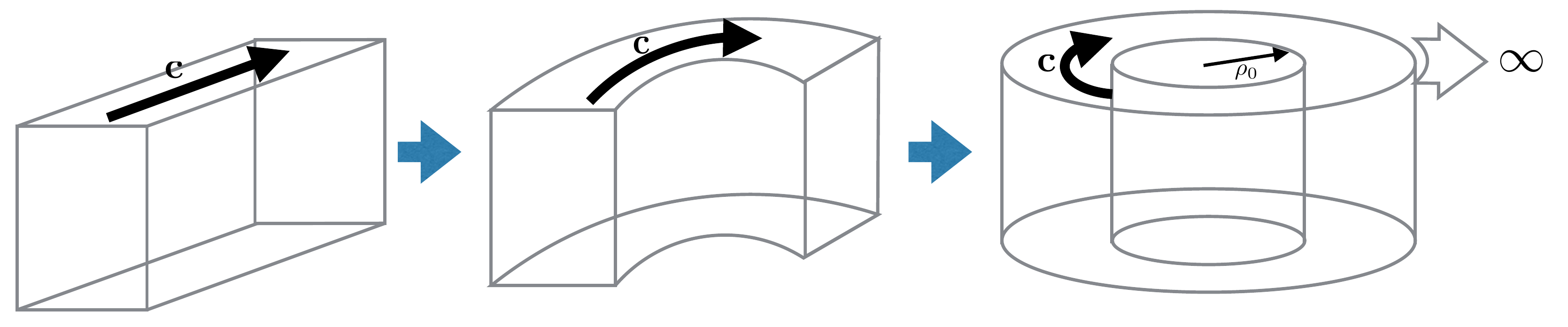}
\caption{Geometry of a Weyl metal sample}\label{CylindericalWM}
\end{figure}

In order to resolve this subtle point, we consider a geometry of the Weyl metal sample as shown in Fig. \ref{CylindericalWM}. We use the differential form since it is independent of the coordinate system and it is easier to calculate the anomaly inflow. The axion term is represented in the following way
\begin{align}
W_{WM}^{Bulk}[\mathcal{A},\mathcal{F}]&=-\int_{\mathcal{M}}\frac{i \theta}{4\pi^2}\mathcal{F}\wedge\mathcal{F}\nonumber \\
&=-\int_{\mathcal{M}}\frac{i\theta}{4\pi^2}d(\mathcal{A}\wedge \mathcal{F})\nonumber \\
&=-\frac{i}{4\pi^2}\int_{\mathcal{M}}[d(\theta\mathcal{A}\wedge \mathcal{F})-d\theta\wedge\mathcal{A}\wedge \mathcal{F}]\nonumber \\
&=\frac{i}{4\pi^2}\int_{\mathcal{M}}d\theta\wedge \mathcal{A}\wedge\mathcal{F} ,
\end{align}
where $\theta\propto c_\mu x^\mu$ is an ``axion" field, $\mathcal{A}=A_\mu dx^\mu$, and $\mathcal{F}=\frac{1}{2}F_{\mu\nu}dx^\mu\wedge dx^\nu$. $\mathcal{M}$ denotes an infinite space, where the Weyl metal sample is embedded. The boundary of the Weyl metal sample is defined by the function $\theta(x)$.

Under the gauge transformation $\mathcal{A}\rightarrow \mathcal{A}+d\eta$, the variation of the effective bulk action is given by
\begin{align}
\delta_\eta W_{WM}^{Bulk}[\mathcal{A},\mathcal{F}]&=\frac{i}{4\pi^2}\int_{\mathcal{M}}d\theta \wedge d\eta \wedge \mathcal{F}\nonumber\\
&=\frac{i}{4\pi^2}\int_{\mathcal{M}}[-d(\eta d\theta\wedge\mathcal{F})+\eta d^2\theta\wedge\mathcal{F}]\nonumber \\
&=\frac{i}{4\pi^2}\int_{\mathcal{M}}\eta d^2\theta\wedge \mathcal{F}
\end{align}
In the cylindrical coordinate $(\rho,\phi,z)$, we can set $\theta(x)=-c\phi\Theta(\rho-\rho_0)$, where $\rho=\rho_0$ represents the boundary of the Weyl-metal sample (Fig. \ref{CylindericalWM}). We emphasize that $\theta(x)$ is not a single-valued function. As a result, $d^2\theta\neq 0$. When $\rho>\rho_0$, $\nabla \theta(x)=-\frac{1}{\rho}c\hat{\phi}$. Therefore, we have $\oint_{C(\rho>\rho_0)} \nabla\theta(x)\cdot dl=-2\pi c$. However, if $\rho<\rho_0$, we have $\oint_{C(\rho<\rho_0)}\nabla\theta(x)\cdot dl=0$. These equations are translated into $\nabla\times\nabla\theta(x)=-\frac{c}{\rho}\delta(\rho-\rho_0)\hat{z}=\hat{z}(\partial_x\partial_y-\partial_y\partial_x)\theta$.

Inserting this equation into the above, we find the anomaly inflow from the bulk state
\begin{align}
\int_{\mathcal{M}}\eta d^2\theta\wedge \mathcal{F}&=\int_{\mathcal{M}}\frac{\eta}{2} (\partial_\mu\partial_\nu\theta)F_{\alpha\beta}dx^\mu\wedge dx^\nu\wedge dx^\alpha \wedge dx^\beta\nonumber \\
&=\int_{\mathcal{M}}d^4x \frac{\eta}{4}\epsilon^{\mu\nu\alpha\beta}(\partial_\mu\partial_\nu-\partial_\nu\partial_\mu)\theta F_{\alpha\beta}\nonumber \\
&=\int d^4x \eta(\partial_x\partial_y-\partial_y\partial_x)\theta F_{zt}\nonumber \\
&=-2\pi c\int dt dz \eta F_{zt} .
\end{align}
The variation of the effective action under the gauge transformation is
\begin{gather}
\delta_{\eta}W_{WM}^{Bulk}[\mathcal{A},\mathcal{F}]=-\frac{ic}{2\pi}\int dtdz\eta(x)F_{zt} . \label{bulkAnomaly}
\end{gather}
Comparing Eq. (\ref{surfaceAnomaly}) with Eq. (\ref{bulkAnomaly}), we confirm the anomaly cancelation
\begin{gather}
\delta_{\eta}W_{WM}^{edge}+\delta_{\eta}W_{WM}^{Bulk}=0 .
\end{gather}

\section{Inversion symmetry-broken Weyl metals}

Since time reversal symmetry is preserved, the net Berry flux should vanish. See Appendix \ref{Appendix:Zero-Berry-flux-IB WM} for the proof of this statement. This implies that there should be an even number of pairs of Weyl points in the inversion symmetry-broken Weyl metal state. Here, we apply the Callan-Harvey mechanism to the inversion symmetry-broken Weyl metal state. We find that a pair of Fermi arcs appear to give rise to a ``quantized" version of either spin or valley Hall effects.

\subsection{An effective minimal model for inversion symmetry-broken Weyl metals}

Following Ref. \cite{Burkov-Kim}, we start from
\begin{gather}
H=\sigma^xs^zk_x-\sigma^yk_y+(-m_1+m_2k_z^2)\sigma^z+\alpha \sigma^x . \label{Minimal_Lattice_Model}
\end{gather}
One can show that the parity and time-reversal transformation operators are given by $P=\sigma^z$ and $T=is^y\mathcal{K}$, respectively, where $\mathcal{K}$ perform the operation of complex conjugation. Then, it is straightforward to see the time reversal symmetry of this effective Hamiltonian. On the other hand, the last term with the coefficient $\alpha$ breaks inversion symmetry. One can find that there are more terms which give rise to breaking the inversion symmetry while preserving the time reversal symmetry: $\sigma^y s^x$, $\sigma^ys^y$, and $\sigma^ys^z$. The first and the second terms result in two nodal rings in momentum space. The third term makes a Weyl point along the $k_y$ direction while the $\alpha$ term causes the Weyl point along the $k_x$ direction.

We start to consider the inversion symmetric case with $\alpha=0$. Since both the time reversal and the inversion symmetry are preserved, two bands must be degenerate. Eigen values are given by
\begin{gather}
E_\pm=\pm\sqrt{k_x^2+k_y^2+(m_2k_z^2-m_1)^2} .
\end{gather}
There are two Dirac points at $(0,0,\pm \sqrt{m_1/m_2})$.

\begin{figure}
\includegraphics[width=0.3\textwidth]{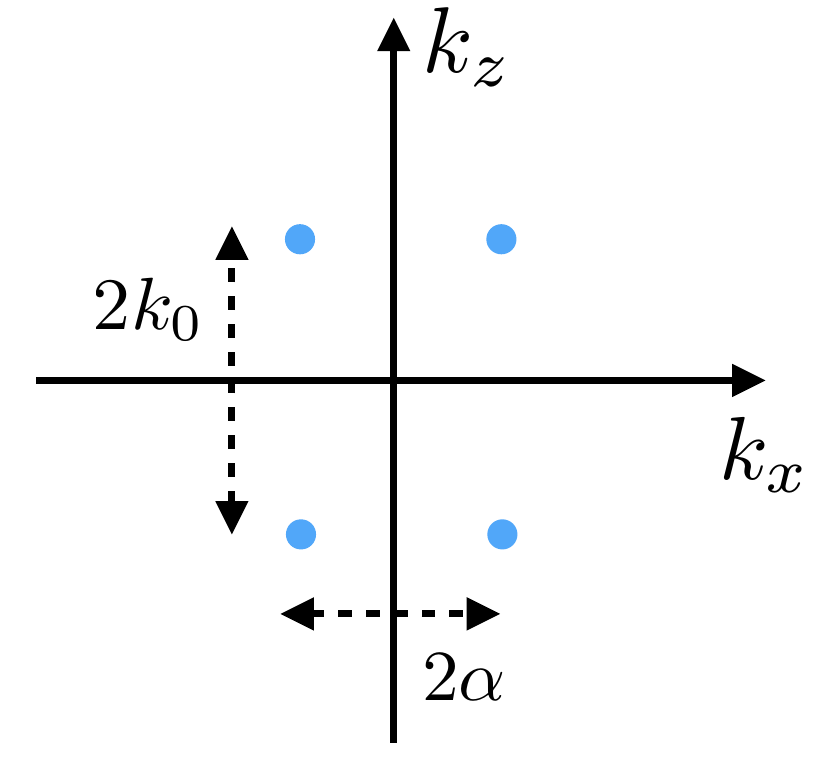}
\caption{Band structure of an effective Hamiltonian Eq. (\ref{Minimal_Lattice_Model}). Four blue dots denote four Weyl points in the $k_y=0$ plane. Here, $k_0=\sqrt{m_1/m_2}$ and $\alpha$ is the strength of inversion symmetry breaking.}\label{WeylPoints}
\end{figure}

Turning on $\alpha$, the band structure evolves into
\begin{gather}
E_{1,\pm}=\pm\sqrt{(k_x-\alpha)^2+k_y^2+(m_2k_z^2-m_1)^2} , \\
E_{2\pm}=\pm\sqrt{(k_x+\alpha)^2+k_y^2+(m_2k_z^2-m_1)^2} .
\end{gather}
Each Dirac point splits into a pair of Weyl points. As a result, we have a double pair of Weyl points at $(\alpha,0,\sqrt{m_1/m_2})$, $(\alpha,0,-\sqrt{m_1/m_2})$, $(-\alpha,0,\sqrt{m_1/m_2})$, and $(-\alpha,0,-\sqrt{m_1/m_2})$ as shown in the Fig. \ref{WeylPoints}. Here, the definition of ``pair" will be clarified below.

\subsection{Low-energy effective Hamiltonian with inversion symmetry breaking}

In order to discuss anomaly cancelation, we write down a low-energy effective Hamiltonian near the double pair of Weyl points shown in Fig. \ref{WeylPoints}. Expanding the momentum near the two Dirac points at $(0,0,\pm\sqrt{m_1/m_2}) \equiv (0,0,\pm k_0) \equiv \pm \mathbf{k}_0$, we obtain
\begin{align}
H^+(\mathbf{k})&=H(\mathbf{k}_0+\delta\mathbf{k})\approx H(\mathbf{k}_0)+\delta\mathbf{k}\cdot\nabla_{k}H(\mathbf{k})|_{\mathbf{k}_0} \nonumber \\
&=\sigma^xs^zk_x-\sigma^yk_y+\alpha\sigma^x+\sigma^z(k_z-k_0) , \\
H^-(\mathbf{k})&=H(-\mathbf{k}_0+\delta\mathbf{k})\approx H(-\mathbf{k}_0)+\delta\mathbf{k}\cdot\nabla_{k}H(\mathbf{k})|_{-\mathbf{k}_0} \nonumber \\
&=\sigma^xs^zk_x-\sigma^yk_y+\alpha\sigma^x-\sigma^z(k_z+k_0) ,
\end{align}
where we set $2\sqrt{m_1m_2}=1$ for simplicity. Then, the original Hamiltonian can be approximated in the low-energy limit as follows
\begin{align}
H(\mathbf{k})\approx H^+(\mathbf{k})f(|\mathbf{k}-\mathbf{k}_0|)+H^-(\mathbf{k})f(|\mathbf{k}+\mathbf{k}_0|) . \label{approximatedk}
\end{align}
Here, the function $f(x)$ is introduced to play the role of a UV cutoff for this low-energy effective Hamiltonian. Accordingly, the Bloch state is represented as
\begin{gather}
|\mathbf{k},\sigma,s\rangle\approx\Big\{\begin{array}{cc}|\mathbf{k},\sigma,s,+\rangle & (\mathbf{k}\sim \mathbf{k}_0) \\ |\mathbf{k},\sigma,s,-\rangle & (\mathbf{k}\sim -\mathbf{k}_0) \end{array} . \label{approximatedk_wavefunction}
\end{gather}

Now, we rewrite this low-energy effective Hamiltonian as one reducible representation in the following way
\begin{align}
\tilde{H}(\mathbf{k})&\approx\Big(\begin{array}{cc}H^+(\mathbf{k}) & 0 \\ 0 & H^-(\mathbf{k})\end{array}\Big)\nonumber \\
&=(\sigma^xs^zk_x-\sigma^yk_y+\alpha\sigma^x-\sigma^zk_0)\otimes \tau^0\nonumber\\
&+\sigma^z\otimes\tau^zk_z .
\end{align}
Accordingly, we have $|\mathbf{k},\sigma,s\rangle \rightarrow |\mathbf{k},\sigma,s,\tau\rangle$, where an additional quantum number is identified with a valley index. $\tilde{H}(\mathbf{k})$ is an eight-band Hamiltonian which occurs from the four-band one Eq. (\ref{Minimal_Lattice_Model}) in the low-energy limit.

The inversion and time-reversal transformation operators are redefined consistently as follows
\begin{gather}
\tilde{P}=\sigma^z\otimes \tau^x,\;\; \tilde{T}=is^y\otimes\tau^x\mathcal{K} .
\end{gather}
See Appendix \ref{Appendix:Inversion-TimeReversal-Operator} for the derivation. It is easy to check out that the time reversal symmetry is preserved for this low-energy effective Hamiltonian, i.e., $\tilde{T}\tilde{H}(\mathbf{k})\tilde{T}^{-1}=\tilde{H}(-\mathbf{k})$ while the inversion symmetry is not respected due to the $\alpha$ term, i.e., $\tilde{P}\tilde{H}(\mathbf{k})\tilde{P}^{-1}\neq \tilde{H}(-\mathbf{k})$.

\begin{widetext}

\subsection{Gamma matrix description}

It is straightforward to write down the low-energy effective Hamiltonian with Gamma matrices. Taking into account the eight-component spinor
\begin{gather}
\Psi=\left(\begin{array}{cccccccc}\phi_{1,1,1}, & \phi_{1,1,-1}, & \phi_{1,-1,1}, & \phi_{1,-1,-1}, & \phi_{-1,1,1}, & \phi_{-1,1,-1}, & \phi_{-1,-1,1}, & \phi_{-1,-1,-1}\end{array}\right)^T ,
\end{gather}
where $\phi_{a,b,c}=\phi_{\tau^z,s^z,\sigma^z}$, we obtain the low-energy effective Hamiltonian
\begin{align}
\mathcal{H}&=\sum_{\mathbf{k}}\Psi^\dagger(\mathbf{k})(\sigma^xs^zk_x-\sigma^yk_y+\alpha\sigma^x+\sigma^z\tau^zk_z-\sigma^zk_0)\Psi(\mathbf{k})\nonumber \\
&=\int d^3 r \Psi^\dagger(r)(\sigma^xs^zi\partial_x-\sigma^yi\partial_y+\sigma^z\tau^zi\partial_z+\sigma^x \alpha-\sigma^zk_0)\Psi(r) .
\end{align}
This gives rise to the following effective action
\begin{align}
S&=\int d^4x\Psi^\dagger(x)[\partial_0+\sigma^xs^zi\partial_x-\sigma^yi\partial_y+\sigma^z\tau^zi\partial_z+\sigma^x\alpha-\sigma^zk_0]\Psi(x)\nonumber \\
&=\int d^4 x \bar{\Psi}(x)[\Gamma^0\partial_0+i\Gamma^1\partial_x+i\Gamma^2\partial_y+i\Gamma^3\partial_z-\alpha \Gamma^1\Gamma^5\tau^z+k_0\Gamma^3\Gamma^5s^z]\Psi(x) , \label{GammaAction}
\end{align}
\end{widetext}
where $\bar{\Psi}(x)=\Psi^\dagger(x)\Gamma^0$ and Gamma matrices are given by
\begin{gather}
\Gamma^0\Gamma^1=\sigma^x s^z,\;\; \Gamma^0\Gamma^2=-\sigma^y,\;\;\Gamma^0\Gamma^3=\sigma^z\tau^z\label{Gamma}\\
\Rightarrow\Gamma^5=i\Gamma^0\Gamma^1\Gamma^2\Gamma^3=-s^z \tau^z ,
\end{gather}
satisfying $\{\Gamma^\mu,\Gamma^\nu\}=2g^{\mu\nu}1_{8\times8}$ with $g^{\mu\nu} = (1,-1,-1,-1) \delta^{\mu\nu}$. We observe that there are two different representations of $\Gamma^\mu$ satisfying Eq. (\ref{Gamma}) with $\{\Gamma^\mu,\Gamma^\nu\}=2g^{\mu\nu}1_{8\times8}$ and $\{\Gamma^\mu,\Gamma^5\}=0$.

\subsubsection{$[\tau^z,\Gamma^\mu]=0$ and $\{s^z,\Gamma^\mu\}=0$}

The first representation for $\Gamma^\mu$ is
\begin{gather}
\Gamma^0=s^x\sigma^x,\;\; \Gamma^1=-is^y,\\
\Gamma^2=-is^x\sigma^z,\;\;\Gamma^3=-is^x\sigma^y\tau^z .
\end{gather}
These eight by eight gamma matrices can be rewritten as a product of four by four gamma matrices and two by two pauli matrices as follows
\begin{gather}
\Gamma^0=\gamma_v^0\tau^0,\; \Gamma^1=\gamma_v^1\tau^0,\; \Gamma^2=\gamma_v^2\tau^0,\; \\
\Gamma^3=\gamma_v^3\tau^z,\; \Gamma^5=\gamma_v^5\tau^z ,
\end{gather}
where
\begin{gather}
\gamma_v^0=s^x\sigma^x,\; \gamma_v^1=-is^y\sigma^0,\; \gamma_v^2=-is^x\sigma^z,\\
\gamma_v^3=-is^x\sigma^y,\; \gamma_v^5=i\gamma_v^0\gamma_v^1\gamma_v^2\gamma_v^3=-s^z\sigma^0 .
\end{gather}
$\gamma_v^\mu$ matrices are four by four matrix, which consist of $\sigma^\mu$ and $s^\mu$, satisfying $\{\gamma_v^\mu,\gamma_v^\nu\}=2g^{\mu\nu}1_{4\times 4}$.

\subsubsection{$\{\tau^z,\Gamma^\mu\}=0$ and $[s^z,\Gamma^\mu]=0$}

The other representation for $\Gamma^\mu$ is
\begin{gather}
\Gamma^0=\sigma^z\tau^x,\;\; \Gamma^1=i\sigma^ys^z\tau^x, \\
\Gamma^2=i\sigma^x\tau^x,\;\; \Gamma^3=-i\tau^y
\end{gather}
These eight by eight gamma matrices can be also rewritten as a product of four by four gamma matrices and two by two pauli matrices as follows
\begin{gather}
\Gamma^0=\gamma_s^0s^0,\; \Gamma^1=\gamma_s^1s^z,\; \Gamma^2=\gamma_s^2s^0,\\
\Gamma^3=\gamma_s^3s^0,\; \Gamma^5=\gamma_s^5s^z ,
\end{gather}
where
\begin{gather}
\gamma_s^0=\sigma^z\tau^x,\; \gamma_s^1=i\sigma^y\tau^x,\; \gamma_s^2=i\sigma^x\tau^x,\\
\gamma_s^3=-i\sigma^0\tau^y,\; \gamma_s^5=i\gamma_s^0\gamma_s^1\gamma_s^2\gamma_s^3=-\sigma^0\tau^z .
\end{gather}
$\gamma_s^\mu$ matrices are four by four matrix, which consist of $\sigma^\mu$ and $\tau^\mu$, satisfying $\{\gamma_s^\mu,\gamma_s^\nu\}=2g^{\mu\nu}1_{4\times 4}$.

\begin{widetext}
As a result, we have two types of low-energy effective Hamiltonians:
\begin{gather}
S_{1st}=\int d^4x\bar{\Psi}(x)\Big[\gamma_v^0\partial_0+i\gamma_v^1\partial_x+i\gamma_v^2\partial_y+i\gamma_v^3\tau^z\partial_z-\alpha\gamma_v^1\gamma_v^5-k_0\gamma_v^3\Big]\Psi(x) , \label{1stAction}\\
S_{2nd}=\int d^4x\bar{\Psi}(x)\Big[\gamma_s^0\partial_0+i\gamma_s^1s^z\partial_x+i\gamma_s^2\partial_y+i\gamma_s^3\partial_z+\alpha\gamma_s^1+k_0\gamma_s^3\gamma_s^5\Big]\Psi(x) . \label{2ndAction}
\end{gather}
\end{widetext}
It is not possible to find the representation satisfying both $[\tau^z,\Gamma^\mu]=0$ and $[s^z,\Gamma^\mu]=0$. If $\Gamma^\mu$ fulfills both conditions, $[\Gamma^\mu,\Gamma^5]=0$ must be satisfied because of $\Gamma^5=-\tau^z s^z$. This is contradictory to the condition of $\{\Gamma^\mu,\Gamma^5\}=0$. It turns out that this property of $\Gamma^\mu$ is related to the Fujikawa's uncertainty principle \cite{Fujikawa-Uncertainty}, which plays an important role in the following discussion.

\subsection{An effective axionic action for inversion symmetry-broken Weyl metals}

Since the total Berry flux is zero for inversion symmetry-broken Weyl metals, the Hall conductivity must vanish. As a result, the conventional effective axionic action does not exist for this Weyl metal state. However, we find other types of effective axionic actions, introducing two kinds of fictitious gauge fields into the effective action: One is a spin gauge field $S_\mu$ and the other is a valley gauge field $V_\mu$, which are coupled with a spin current $j_s^\mu=\bar{\Psi}\Gamma^\mu s^z\Psi$ and a valley current $j_v^{\mu}=\bar{\Psi}\Gamma^\mu \tau^z \Psi$, respectively. Both spin and valley currents are Noether currents, involved with the symmetry under $\Psi\rightarrow e^{is^z\theta}\Psi$ and $\Psi\rightarrow e^{i\tau^z\theta}\Psi$, respectively.

We start from the low-energy effective action with both spin and valley gauge fields
\begin{widetext}
\begin{gather}
S=\int d^4x \bar{\Psi}(x)\Big[i\Gamma^\mu(\partial_\mu+iA_\mu+is^zS_\mu+i\tau^zV_\mu)-\alpha\Gamma^1\Gamma^5\tau^z+k_0\Gamma^3\Gamma^5s^z\Big]\Psi(x) ,
\end{gather}
\end{widetext}
where $\mu=1,2,3,4$ and $\Gamma^4=-i\Gamma^0$. It turns out that both spin and valley currents can not be conserved simultaneously when quantum corrections are taken into account. In other words, one of both symmetries related to either spin or valley current should be anomalous in the quantum level. The problem on which symmetry becomes anomalous should be determined by the UV condition. The UV condition fixes the possible representation for the low-energy effective field theory. Physically, this determines the formation of a pair of Fermi arcs.

\subsubsection{$[\tau^z,\Gamma^\mu]=0$ and $\{s^z,\Gamma^\mu\}=0$}

Action in the 1st representation Eq.($\ref{1stAction}$) is symmetric under the following three kinds of transformations
\begin{gather}
\Psi\rightarrow e^{i\alpha(x)}\Psi,\; \bar{\Psi}\rightarrow \bar{\Psi}e^{-i\alpha(x)} , \\
\Psi\rightarrow e^{i\tau^z\beta(x)}\Psi,\; \bar{\Psi}\rightarrow \bar{\Psi}e^{-i\tau^z\beta(x)} , \\
\Psi\rightarrow e^{is^z\eta(x)}\Psi,\; \bar{\Psi}\rightarrow \bar{\Psi}e^{is^z\eta(x)}\; (\because s^z=-\gamma_v^5) .
\end{gather}
The first, second, and third transformations are related to the charge, valley, and spin current, respectively. We note that the third transformation related to the spin current is the chiral transformation in terms of the $\gamma_v$ matrix. Therefore, this low-energy effective action is not invariant under the third transformation when quantum corrections are included. In mathematical terms, the integral measure of the partition function is not invariant under the third transformation. As a result, the spin current is not a conserved current. Resorting to the Fujikawa's method, one can obtain an effective axion term as we did in the case of time-reversal symmetry-broken Weyl metals. Detailed calculations are shown in Appendix \ref{Appendix:Axionic-Action-IB-Weyl}. Here, we quote the result only
\begin{gather}
S_{eff}^{1st} \equiv S_{eff}^v=-\frac{i}{4\pi^2}\int d^4x \alpha x^1\epsilon^{\mu\nu\alpha\beta}F_{v,\mu\nu}F_{\alpha\beta}
\end{gather}
where $F_{v,\mu\nu}=\partial_\mu V_\nu-\partial_\nu V_\mu$ is the field strength tensor, given by the valley gauge field $V_\mu$. We point out that the coefficient of this effective action is four times larger than that of the time-reversal symmetry-broken Weyl metal. It turns out that this enhancement plays an essential role in the anomaly cancelation.

Following the same method as the case of time-reversal symmetry-broken Weyl metals, it is straightforward to find the valley Hall current from this axion term. Performing the integration by part as follows
\begin{align}
S_{eff}^v&=-\int d^4x\frac{i\alpha x^1}{4\pi^2}\epsilon^{\mu\nu\alpha\beta}F_{v,\mu\nu}F_{\alpha\beta}\equiv\int_M\frac{i\theta_v(x^1)}{\pi^2}\mathcal{F}_v\wedge\mathcal{F}\nonumber \\
&=\int_M\frac{i}{\pi^2}[d(\theta_v\mathcal{V}\wedge\mathcal{F})-d\theta_v\wedge\mathcal{V}\wedge\mathcal{F}]\nonumber \\
&=- \frac{i}{\pi^2}\int_Md\theta_v\wedge\mathcal{V}\wedge\mathcal{F} \nonumber \\
&=-\frac{i}{2\pi^2}\int d^4x \epsilon^{\alpha\beta\mu\nu}\partial_\alpha \theta_vV_{\beta}F_{\mu\nu} ,
\end{align}
we obtain
\begin{align}
j_{v}^{\mu}&=\frac{\delta S_{eff}^v}{\delta V_{\mu}}=\frac{i}{2\pi^2}\epsilon^{\mu\nu\alpha\beta}(\partial_\nu\theta_v)F_{\alpha\beta}\nonumber \\
&=-\frac{i}{2\pi^2}\epsilon^{\mu 1\eta\xi}\alpha F_{\eta\xi} .
\end{align}
%
%

Since this current is evaluated in the Euclidean signature, we have to change it into the Lorentzian signature; $(v_L^1,v_L^2,v_L^3,v_L^0)=(v_E^1,v_E^2,v_E^3,iv_E^4)$. Then, we have
\begin{gather}
j_{L,v}^0=\frac{\alpha}{\pi^2}F_{23} , \\
j_{L,v}^k=\frac{\alpha}{2\pi^2}\epsilon^{k1\eta\xi}F_{\eta\xi} .
\end{gather}
This valley Hall current may be regarded as an anomaly inflow from the bulk to the pair of Fermi arcs. Performing essentially the same task as the case of the time-reversal symmetry-broken Weyl metal, we find that the pair of Fermi arcs is given in Fig. \ref{ValleyFermiArc}.

\begin{figure}
\includegraphics[width=0.25\textwidth]{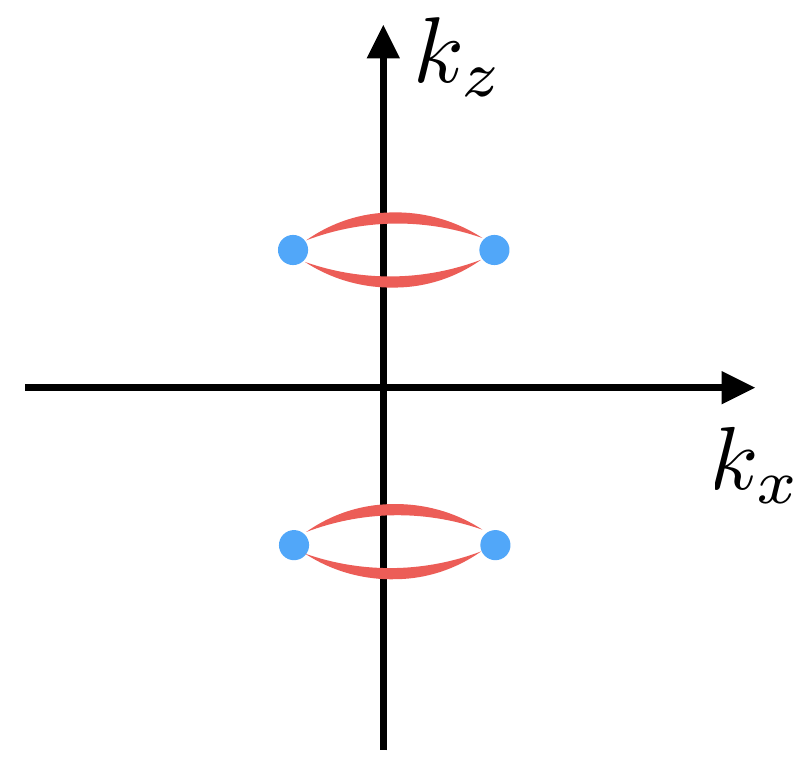}
\caption{A pair of Fermi arcs (red lines) for the case of $[\tau^z,\Gamma^\mu]=0$ and $\{s^z,\Gamma^\mu\}=0$} \label{ValleyFermiArc}
\end{figure}

\subsubsection{$\{\tau^z,\Gamma^\mu\}=0$ and $[s^z,\Gamma^\mu]= 0$}

The low-energy effective action Eq.(\ref{2ndAction}) is symmetric under the following three types of transformations as the case of the 1st representation
\begin{gather}
\Psi\rightarrow e^{i\alpha(x)}\Psi,\; \bar{\Psi}\rightarrow \bar{\Psi}e^{-i\alpha(x)} , \\
\Psi\rightarrow e^{i\tau^z\beta(x)}\Psi,\; \bar{\Psi}\rightarrow \bar{\Psi}e^{i\tau^z\beta(x)}\;(\because \tau^z=-\gamma_s^5) , \\
\Psi\rightarrow e^{is^z\eta(x)}\Psi,\; \bar{\Psi}\rightarrow \bar{\Psi}e^{-is^z\eta(x)} .
\end{gather}
In this representation the second transformation related to the valley current is the chiral transformation in terms of the $\gamma_s^5$ matrix. Therefore, the valley current is not conserved because of the chiral anomaly. The corresponding axionic effective action derived from the chiral rotation is given by
\begin{gather}
S_{eff}^{2nd}\equiv S_{eff}^s=\frac{i}{4\pi^2}\int d^4x k_0x^3 \epsilon^{\mu\nu\alpha\beta}F_{s,\mu\nu}F_{\alpha\beta}
\end{gather}
where $F_{s,\mu\nu}=\partial_\mu S_\nu-\partial_\nu S_\mu$ is the field strength of the spin gauge field $S_\mu$. We refer all details to Appendix \ref{Appendix:Axionic-Action-IB-Weyl}.

It is straightforward to find the spin Hall current from this effective action, taking into account the integration by part
\begin{align}
S_{eff}^s&=\int d^4x\frac{ik_0 x^3}{4\pi^2}\epsilon^{\mu\nu\alpha\beta}F_{s,\mu\nu}F_{\alpha\beta}=\int_M\frac{i\theta_s(x^1)}{\pi^2}\mathcal{F}_s\wedge\mathcal{F}\nonumber \\
&=\int_M\frac{i}{\pi^2}[d(\theta_s\mathcal{S}\wedge\mathcal{F})-d\theta_s\wedge\mathcal{S}\wedge\mathcal{F}]\nonumber \\
&=- \frac{i}{\pi^2}\int_Md\theta_s\wedge\mathcal{S}\wedge\mathcal{F}\nonumber \\
&=-\frac{i}{2\pi^2}\int d^4x \epsilon^{\alpha\beta\mu\nu}\partial_\alpha \theta_sS_{\beta}F_{\mu\nu} ,
\end{align}
and resulting in
\begin{align}
j_{s}^{\mu}&=\frac{\delta S_{eff}^v}{\delta S_{\mu}}=\frac{i}{2\pi^2}\epsilon^{\mu\nu\alpha\beta}(\partial_\nu\theta_s)F_{\alpha\beta}\nonumber \\
&=\frac{i}{2\pi^2}\epsilon^{\mu 3\eta\xi}k_0 F_{\eta\xi} .
\end{align}
In the Lorentizan signature we have
\begin{gather}
j_{L,s}^{0}=-\frac{k_0}{\pi^2}F_{12} , \\
j_{L,s}^{k}=-\frac{k_0}{2\pi^2}\epsilon^{k3\eta\xi}F_{\eta\xi} .
\end{gather}

\begin{figure}
\includegraphics[width=0.25\textwidth]{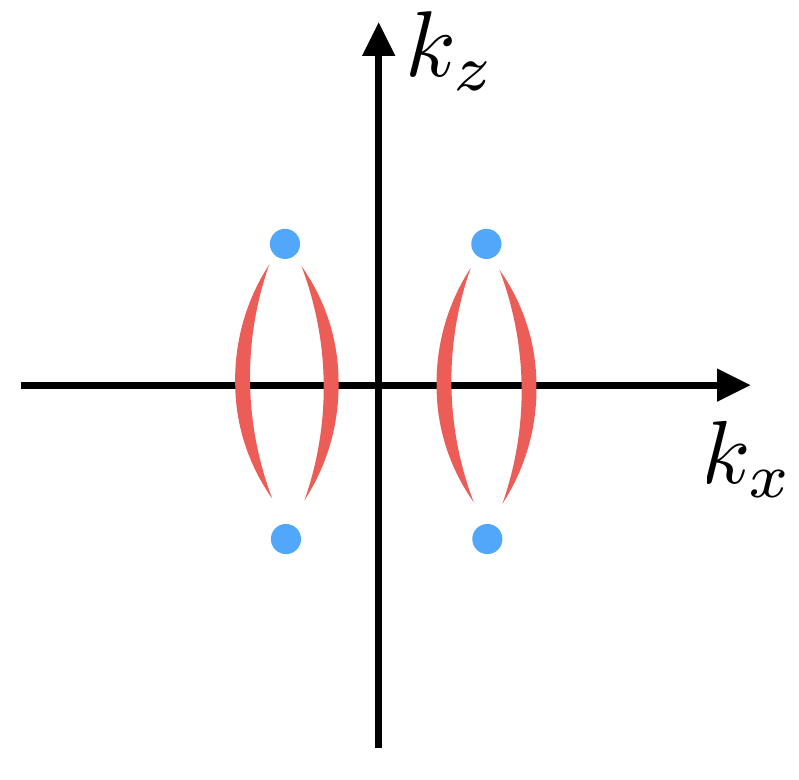}
\caption{A pair of Fermi arcs (red lines) for the case of $\{\tau^z,\Gamma^\mu\}=0$ and $[s^z,\Gamma^\mu]= 0$}\label{SpinFermiArc}
\end{figure}

This spin Hall current may be also regarded as an anomaly inflow from the bulk to the pair of Fermi arcs. Performing essentially the same task as the case of the time-reversal symmetry-broken Weyl metal, we find that the pair of Fermi arcs is given in Fig. \ref{SpinFermiArc}.

So far, we discussed that two different representations give two different physical situations. Then, what is the right physical picture? We cannot answer which is correct within only the low energy effective Hamiltonian. We need more information which should be introduced from the UV structure of the dispersion relation. However, the analysis based on the low energy effective action tells us that only one of these two different choices exists. These two representations take two different regularization schemes: The first representation or regularization scheme preserves the charge symmetry and the valley symmetry, which cannot but break the spin symmetry while the second representation or regularization scheme preserves the charge symmetry and the spin symmetry, which cannot but break the valley symmetry. One important thing is that there is no regularization scheme which preserves all the symmetry involved with charge, spin, and valley simultaneously. Although we have shown this aspect with a specific Hamiltonian, this argument can be generalized. See Appendix \ref{Appendix:general proof} for more general discussions on this point.

Additionally, effective axionic actions of $S_{eff}^v[A_\mu,V_\mu]$ and $S_{eff}^s[A_\mu,S_\mu]$ have anomaly with respect to the valley gauge transformation of $V_\mu\rightarrow V_\mu+\partial_\mu \eta$ and the spin gauge transformation of $S_\mu\rightarrow S_\mu+\partial_\mu\eta$, respectively. We find
\begin{gather}
\delta_\eta S_{eff}^v[\mathcal{A},\mathcal{V}]=\frac{2i\alpha}{\pi}\int dt dx \eta F_{xt}\\
\delta_\eta S_{eff}^s[\mathcal{A},\mathcal{S}]=-\frac{2ik^0}{\pi}\int dtdy \eta F_{zt}
\end{gather}
Of course, this anomaly inflow should be canceled by the anomaly of a pair of Fermi arcs, which is nothing but the Callan-Harvey mechanism.

\subsection{Callan-Harvey mechanism in inversion symmetry-broken Weyl metals}

\subsubsection{Valley Hall current}

Following the case of time-reversal symmetry-broken Weyl metals and setting $\alpha\rightarrow \alpha\theta(z)$, one can find the edge state localized near the boundary from the low-energy effective Hamiltonian. We show all detailed calculations in Appendix \ref{Appendix:Boundary-Mode-IB-Weyl}.

An effective surface Hamiltonian in the $|\tau^z,k_x,k_y\rangle$ basis is given by
\begin{align}
H&=\sum_{\tilde{k}_z}\sum_{k_y}\Psi^\dagger(k_x,k_y)\left(\begin{array}{cc} k_y &0 \\ 0 & -k_y\end{array}\right)\Psi(k_x,k_y)\nonumber \\
&=\sum_{\tilde{k}_z}\int dy \Psi_{k_x}^\dagger(y)\left(\begin{array}{cc}-i\partial_y & 0 \\ 0 & i\partial_y\end{array}\right)\Psi_{k_x}(y)\nonumber \\
&=\sum_{\tilde{k}_z}\int dy \Psi_{k_x}^\dagger(y)(-i\partial_y \tau^3)\Psi_{k_x}(y) ,
\end{align}
where $\tilde{k}_z\rightarrow-\sqrt{\alpha^2-m^2}<k_x<\sqrt{\alpha^2-m^2}$. Then, the corresponding effective surface action is
\begin{align}
S&=\sum_{\tilde{k}_z}\int d\tau \int dy\Psi^\dagger_{k_x}(y)\Big(\partial_\tau-i\tau^3\partial_y\Big)\Psi_{k_x}(y)\nonumber \\
&=\sum_{\tilde{k}_z}\int d\tau \int dy\bar{\Psi}_{k_x}(y)\Big(\gamma^0\partial_\tau+i\gamma^1\partial_y\Big)\Psi_{k_x}(y) ,
\end{align}
where $\gamma^0=\tau^1$, $\gamma^1=i\tau^2$, and $\bar{\gamma}=\gamma^0\gamma^1=-\tau^3$. Setting $\gamma^2=-i\gamma^0$, we have
\begin{gather}
S=\sum_{-\sqrt{\alpha^2-m^2}<k_x<\sqrt{\alpha^2-m^2}}\int d^2x \bar{\Psi}_{k_x}(y)i\gamma^\mu\partial_\mu\Psi_{k_x}(y) ,
\end{gather}
where $\mu=1,2$, $\partial_0=\partial_2$, and $\{\gamma^\mu,\gamma^\nu\}=-\delta^{\mu\nu}$.

In order to show the anomaly cancelation, we introduce both charge and valley gauge fields to the above boundary action
\begin{widetext}
\begin{gather}
S=\sum_{-\sqrt{\alpha^2-m^2}<k_x<\sqrt{\alpha^2-m^2}}\int d^2 x\bar{\Psi}_{k_x}(y)i\gamma^\mu(\partial_\mu+iA_\mu+iV_\mu\bar{\gamma})\Psi_{k_x}(y) .
\end{gather}
\end{widetext}
The surface valley current is given by
\begin{align}
j_v^{\mu}&\equiv\frac{\delta W[A,V]}{\delta V_\mu} \nonumber \\
&= -\sum_{-\sqrt{\alpha^2-m^2}<k_x<\sqrt{\alpha^2-m^2}}\langle \bar{\Psi}_{k_x}\gamma^\mu\bar{\gamma}\Psi_{k_x}\rangle ,
\end{align}
where $W[A,V]$ is an effective free energy defined by $Z=e^{-W[A,V]}=\int \mathcal{D}\bar{\Psi}\mathcal{D}\Psi e^{-S[\bar{\Psi},\Psi,A,V]}$.

This boundary effective action is invariant under the following valley gauge transformation
\begin{gather}
\Psi_{k_x}\rightarrow e^{i\bar{\gamma}\theta}\Psi_{k_x},\; \bar{\Psi}_{k_x}\rightarrow \bar{\Psi}_{k_x}e^{i\bar{\gamma}\theta},\; V_\mu\rightarrow V_\mu+\partial_\mu \theta.
\end{gather}
However, we find that the expectation value for the valley current $j_v^\mu$ becomes anomalous in the one-loop order. All details are shown in Appendix \ref{Appendix:1-loop-IB-Weyl}. Here, we quote the result only
\begin{gather}
\partial_\mu \langle j_{reg}^\mu(x)\rangle =\frac{2i\alpha}{\pi}\epsilon^{\mu\nu}\partial_\mu A_\nu (x)=\frac{i\alpha}{\pi}\epsilon^{\mu\nu}F_{\mu\nu}(x) .
\end{gather}
This anomaly equation implies
\begin{gather}
\delta_\eta W_{WM,valley}^{edge}=-\frac{i\alpha}{\pi}\int d^2x \eta(x)\epsilon^{\mu\nu}F_{\mu\nu}(x) .
\end{gather}
$\delta_\eta W_{WM,valley}^{edge}$ is canceled exactly by $\delta_{\eta} S_{eff}^v$, which is nothing but the Callan-Harvey mechanism.

\subsubsection{Spin Hall current}

All processes are completely the same as those of the case of the valley Hall current. See Appendix \ref{Appendix:Boundary-Mode-IB-Weyl} and Appendix \ref{Appendix:1-loop-IB-Weyl} for more details. The variation of the effective surface action with respect to the change of the spin gauge field is
\begin{gather}
\delta_{\eta}W_{WM,spin}^{edge}=\frac{ik_0}{\pi}\int d^2x \eta(x)\epsilon^{\mu\nu}F_{\mu\nu}(x) .
\end{gather}
$\delta_\eta W_{WM,spin}^{edge}$ is also canceled perfectly by $\delta_\eta S_{eff}^s$.

\section{Conclusion}

In conclusion, we applied the Callan-Harvey mechanism to the case of inversion symmetry-broken Weyl metals, and found either the spin Hall effect or the valley Hall effect, depending on the UV condition. Turning on charge, spin, and valley U(1) gauge fields, we derived two types of axion terms from two kinds of low-energy effective actions, based on the Fujikawa's method. This explicit demonstration clarifies the anomaly inflow in either spin or valley currents from the Weyl metal bulk to the surface state. Solving Weyl metal equations with a surface boundary condition, we found normalizable surface zero modes, which consist of a chiral pair of Fermi arcs. Constructing the corresponding effective surface action and calculating both spin and valley surface currents, we found the gauge anomaly involved with either spin or valley U(1) gauge fields. We proved explicitly that this spin-gauge or valley-gauge anomaly at the surface is canceled exactly by the anomaly inflow from the bulk action. We would like to emphasize that our demonstration is the first concrete calculation for inversion symmetry-broken Weyl metals although it is certainly expected in a conceptual point of view.

We believe that our explicit demonstration casts various interesting questions, involved with generalizations of the present ideal case. As far as we know, the role of disorder scattering has never been discussed clearly, particularly, in the view of anomaly cancelation. Disorder scattering gives rise to mixing between each Fermi point in the Fermi arc state, expected to spoil the present simple calculation at least when disorder strength exceeds a critical value. The role of electron correlations in the anomaly cancelation would be more important and difficult. Recently, a topological Fermi-liquid theory has been proposed to describe anomalous transport phenomena in time-reversal symmetry-broken Weyl metals, where the concept of Landau's Fermi-liquid theory is generalized to incorporate both the Berry curvature and the chiral anomaly \cite{TFL1,TFL2}. However, the issue on anomaly cancelation has not been discussed within such a topological Fermi-liquid theory. When inversion symmetry is broken instead of time reversal symmetry, the situation would be much more complicated. Not only the spin current but also the valley current should be taken into account. This is somewhat analogous to the relationship between the integer quantum Hall phase and the quantum spin-Hall state in two dimensions.

\section{Acknowledgement}

This study was supported by the Ministry of Education, Science, and Technology (No. NRF-2015R1C1A1A01051629 and No. 2011-0030046) of the National Research Foundation of Korea (NRF). I. J. was also supported by Global Ph. D. Fellowship of the National Research Foundation of Korea (NRF). We would like to appreciate fruitful discussions in the APCTP Focus program ``Lecture series on Beyond Landau Fermi liquid and BCS superconductivity near quantum criticality" in 2016.

\widetext

\appendix

\section{Derivation of an axionic action for time-reversal symmetry-broken Weyl metals} \label{Axionic-action-TRB WM}

\subsection{Chiral transformation}

We introduce the chiral rotation as follows
\begin{align}
\Psi(x)\rightarrow e^{i\alpha(x)\gamma^5}\Psi(x),\;\; \bar{\Psi}(x)\rightarrow \bar{\Psi}(x)e^{i\alpha(x)\gamma^5} .
\end{align}
Under this chiral transformation, the effective action Eq. (\ref{Weyl_Metal_TRSB}) for a Weyl metal state changes as
\begin{align}
S_{WM}&\rightarrow\int d^4x \bar{\Psi}(x)i\gamma^\mu [\partial_\mu +i A_\mu +i(c_\mu+ds \partial_\mu \theta(x))\gamma^5]\Psi(x).
\end{align}
Here, we set $\alpha(x)=ds \theta(x)$. Multiple steps of chiral rotations result in
\begin{align}
&S_{WM}\rightarrow \int d^4x \bar{\Psi}(x)i\gamma^\mu [\partial_\mu +i A_\mu +i(c_\mu+s \partial_\mu \theta(x))\gamma^5]\Psi(x)\equiv\int d^4x \bar{\Psi}(x)i\slashed{D}^{(s)}\Psi(x) ,
\end{align}
where
\begin{align}
&\slashed{D}^{(s)}\equiv \gamma^\mu(\partial_\mu +iA_\mu+i(c_\mu+s\partial_\mu \theta(x)) \gamma^5) , \\
&\slashed{D}^{(s)\dagger}\equiv\gamma^\mu(\partial_\mu+iA_\mu-i(c_\mu+s\partial_\mu \theta(x))\gamma^5) .
\end{align}

Since $\slashed{D}^{(s)}$ is not Hermitian because of the chiral gauge field, we choose a basis which differs from the conventional case of the chiral anomaly \cite{Anomalies in QFT}
\begin{align}
\Psi(x)=\sum_{n}a_n \varphi^{(s)}_{n}(x),\;\; \bar{\Psi}(x)=\sum_{n}\phi^{(s)\dagger}_{n}(x)\bar{b}_{n} ,
\end{align}
where the eigen vectors $\varphi^{(s)}_{n}(x)$ and $\phi^{(s)\dagger}_{n}(x)$ are determined by
\begin{gather}
\slashed{D}^{(s)\dagger}\slashed{D}^{(s)}\varphi_n^{(s)}=\lambda_n^2 \varphi_n^{(s)},\;\;\slashed{D}^{(s)}\slashed{D}^{(s)\dagger}\phi_n^{(s)}=\lambda_n^2 \phi_n^{(s)} , \\
\slashed{D}^{(s)}\varphi_n^{(s)}=\lambda_n\phi_n^{(s)},\;\; \slashed{D}^{(s)\dagger}\phi_n^{(s)}(x)=\lambda_n \varphi_n^{(s)} ,
\end{gather}
%
%
where $\lambda_n$ is an eigen value. Then, the path-integral measure is
\begin{gather}
\mathcal{D}\bar{\Psi}(x)\mathcal{D}\Psi(x)=[det U]^{-1}\prod_n d\bar{b}_n da_n ,
\end{gather}
where $[U^{-1}]_{nm} = \phi^{(s)\dagger}_{n}(x) \varphi^{(s)}_{m}(x)$.

Now, we can see how the integral measure changes under the chiral transformation. Since the wave function changes in the following way
\begin{align}
&\Psi'(x)=e^{ids\theta(x)\gamma^5}\Psi(x),\; \bar{\Psi}'(x)=\bar{\Psi}(x)e^{ids\theta(x)\gamma^5} \nonumber \\
&\Rightarrow\Big\{\begin{array}{c}\sum_{n}a_n'\varphi_n(x)=\sum_n e^{ids \theta(x)\gamma^5}a_n\varphi_n(x), \\ \sum_{n}\phi^{(s)\dagger}_n\bar{b}'_n=\sum_{n}\phi^{(s)\dagger}_n(x)\bar{b}_ne^{ids\theta(x)\gamma^5} \end{array} ,
\end{align}
we obtain
\begin{gather}
a'_n=\sum_{m}C_{nm}a_m,\;\; \bar{b}'_n=\sum_{m}D_{nm}\bar{b}_m , \\
C_{nm}=\int d^dx \varphi^{(s)\dagger}_n(x)e^{ids\theta(x)\gamma^5}\varphi^{(s)}_{m}(x),\\ D_{nm}=\int d^d x \phi^{(s)\dagger}_m e^{ids\theta(x)\gamma^5}\phi^{(s)}_n(x) .
\end{gather}
As a result, the integral measure is given by
\begin{gather}
\mathcal{D}\bar{\Psi}'(x)\mathcal{D}\Psi'(x)=[detU]^{-1}\prod_n d\bar{b}'_nda'_n=[detU]^{-1}[det\; C]^{-1}[det\; D]^{-1}\prod_n d\bar{b}_nda_n=[det\; C]^{-1}[det\; D]^{-1}\mathcal{D}\bar{\Psi}(x)\mathcal{D}\Psi(x)
\end{gather}
under the chiral transformation, where
\begin{align}
[det\; C]^{-1}&=\exp\Big[-ids \int d^dx\theta(x) \sum_n \varphi_n^{(s)\dagger}(x)\gamma^5\varphi^{(s)}_n(x)\Big] , \nonumber\\
[det\; D]^{-1}&=\exp\Big[-ids \int d^dx \theta(x)\sum_n \phi^{(s)\dagger}_n(x)\gamma^5\phi_n^{(s)}(x)\Big] .
\end{align}

Finally, the partition function changes as follows
\begin{align}
&Z=\int \mathcal{D}\bar{\Psi}(x)\mathcal{D}\Psi(x)e^{-S_{WM}}\nonumber \\
&\rightarrow \int \mathcal{D}\bar{\Psi}(x)\mathcal{D}\Psi(x)\exp\Big[-\int d^dx \Big\{\bar{\Psi}(x)i\slashed{D}^{(s)}\Psi(x)+\int_0^s ds \; \theta(x) i\Big(\sum_{n}\varphi_n^{(s)\dagger}(x)\gamma^5\varphi^{(s)}(x)+\sum_n \phi^{(s)\dagger}_n(x)\gamma^5\phi^{(s)}_n(x)\Big)\Big\}\Big]\nonumber\\
&\equiv \int \mathcal{D}\bar{\Psi}(x)\mathcal{D}\Psi(x)\exp\Big[-S_{WM}^{(s)}\Big] .
\end{align}

\subsection{Regularization}

In order to calculate the part which changes under the chiral transformation, we follow the standard way of regularization \cite{Anomalies in QFT}, given by
\begin{align}
\sum_n[\varphi_{n}^{(s)\dagger}(x)\gamma^5\varphi_n^{(s)}+\phi^{(s)\dagger}\gamma^5 \phi^{(s)}_n]&=\lim_{M\rightarrow\infty}\sum_n[\varphi_{n}^{(s)\dagger}\gamma^5\varphi_n^{(s)}+\phi_n^{(s)\dagger}\gamma^5 \phi^{(s)}_n]e^{-\frac{\lambda_n^2}{M^2}}\nonumber\\
&=\lim_{M\rightarrow\infty}\sum_n [\varphi_{n}^{(s)\dagger}\gamma^5e^{-\frac{\slashed{D}^{(s)\dagger}\slashed{D}^{(s)}}{M^2}}\varphi_n^{(s)}+\phi_n^{(s)\dagger}\gamma^5 e^{-\frac{\slashed{D}^{(s)}\slashed{D}^{(s)\dagger}}{M^2}} \phi^{(s)}_n] .
\end{align}

One can show
%
%
\begin{align}
&\sum_{n}\phi_{n}^{(s)\dagger}(x)\gamma^5 e^{-\frac{\slashed{D}^{(s)}\slashed{D}^{(s)\dagger}}{M^2}}\phi_n^{(s)}(x)=\int \frac{d^4k}{(2\pi)^4}e^{-ik\cdot x}tr\Big[\gamma^5e^{-\frac{\slashed{D}^{(s)}\slashed{D}^{(s)\dagger}}{M^2}}\Big]e^{ik\cdot x} .
\end{align}
As a result, we obtain
\begin{align}
\sum_n[\varphi_n^{(s)\dagger}(x)\gamma^5 \varphi_n^{(s)}(x)+\phi^{(s)\dagger}_n\gamma^5\phi_n^{(s)}(x)]=\lim_{M\rightarrow \infty}\int \frac{d^4 k}{(2\pi)^4}e^{-ik\cdot x}tr\Big[\gamma^5\Big( e^{-\frac{\slashed{D}^{(s)\dagger}\slashed{D}^{(s)}}{M^2}}+e^{-\frac{\slashed{D}^{(s)}\slashed{D}^{(s)\dagger}}{M^2}}\Big)\Big]e^{ik\cdot x} .
\end{align}

\subsection{Chirality Splitting}

In order to perform the momentum integration in the above expression, we consider chirality splitting given by
\begin{align}
&\slashed{D}^{(s)}=(\slashed{\partial}+i\slashed{A}_+^{(s)})P_++(\slashed{\partial}+i\slashed{A}_-^{(s)})P_-\equiv \slashed{D}^{(s)}_+P_++\slashed{D}^{(s)}_-P_- , \\
&\slashed{D}^{(s)\dagger}=\slashed{D}^{(s)}_+P_-+\slashed{D}^{(s)}_-P_+ ,
\end{align}
where
\begin{gather}
A_{\mu+}^{(s)}\equiv A_\mu+c_\mu+s\partial_\mu\theta(x) , \\
A_{\mu -}^{(s)}\equiv A_\mu-(c_\mu+s\partial_\mu\theta(x)) , \\
P_{\pm}=\frac{1}{2}(1\pm \gamma^5) , \\
\slashed{D}_{\pm}^{(s)}=\slashed{\partial}+i\slashed{A}_{\pm}^{(s)} .
\end{gather}
Then, we obtain
\begin{gather}
\slashed{D}^{(s)}\slashed{D}^{(s)\dagger}=(\slashed{D}_{+}^{(s)})^2P_-+(\slashed{D}_{-}^{(s)})^2 P_+,\;\; \slashed{D}^{(s)\dagger}\slashed{D}^{(s)}=(\slashed{D}_{+}^{(s)})^2P_++(\slashed{D}_{-}^{(s)})^2 P_- ,
\end{gather}
giving rise to
\begin{gather}
e^{-\frac{\slashed{D}^{(s)\dagger}\slashed{D}^{(s)}}{M^2}}=P_+ e^{-\frac{(\slashed{D}_+^{(s)})^2}{M^2}}+P_- e^{-\frac{(\slashed{D}_-^{(s)})^2}{M^2}} , \\
e^{-\frac{\slashed{D}^{(s)}\slashed{D}^{(s)\dagger}}{M^2}}=P_- e^{-\frac{(\slashed{D}_+^{(s)})^2}{M^2}}+P_+ e^{-\frac{(\slashed{D}_-^{(s)})^2}{M^2}} .
\end{gather}

Now, it is straightforward to perform the momentum integration in the following way
\begin{align}
& \sum_n[\varphi_n^{(s)\dagger}(x)\gamma^5 \varphi_n^{(s)}(x)+\phi^{(s)\dagger}_n\gamma^5\phi_n^{(s)}(x)]=\lim_{M\rightarrow \infty}\int \frac{d^4 k}{(2\pi)^4}e^{-ik\cdot x}tr\Big[\gamma^5\Big( e^{-\frac{(\slashed{D}_+^{(s)})^2}{M^2}}+e^{-\frac{(\slashed{D}^{(s)}_-)^2}{M^2}}\Big)\Big]e^{ik\cdot x} \nonumber\\
&=\lim_{M\rightarrow \infty}\int\frac{d^4k}{(2\pi)^4}tr\Big[\gamma^5 \Big(e^{-\frac{-(D_{+\mu}^{(s)}+ik_\mu)^2+\frac{i}{4}[\gamma^\mu,\gamma^\nu]F_{+,\mu\nu}^{(s)}}{M^2}}+e^{-\frac{-(D_{-\mu}^{(s)}+ik_\mu)^2+\frac{i}{4}[\gamma^\mu,\gamma^\nu]F_{-,\mu\nu}^{(s)}}{M^2}}\Big)\Big]\nonumber\\
%
%
&=\lim_{M\rightarrow \infty}\int\frac{d^4k}{(2\pi)^4}M^4e^{-k_\mu^2}tr\Big[\gamma^5\Big(-\frac{1}{8M^4}\gamma^\mu\gamma^\nu\gamma^\alpha\gamma^\beta F_{+,\mu\nu}^{(s)}F_{+,\alpha\beta}^{(s)}-\frac{1}{8M^4}\gamma^\mu\gamma^\nu\gamma^\alpha\gamma^\beta F_{-,\mu\nu}^{(s)}F_{-,\alpha\beta}^{(s)}\Big)\Big]\nonumber\\
&=\int \frac{d^4k}{(2\pi)^4} e^{-k_\mu^2} \frac{1}{2} \epsilon^{\mu\nu\alpha\beta} [F_{+,\mu\nu}^{(s)}F_{+,\alpha\beta}^{(s)}+F_{-,\mu\nu}^{(s)}F_{-,\alpha\beta}^{(s)}] = \frac{1}{32\pi^2} \epsilon^{\mu\nu\alpha\beta} [F_{+,\mu\nu}^{(s)}F_{+,\alpha\beta}^{(s)}+F_{-,\mu\nu}^{(s)}F_{-,\alpha\beta}^{(s)}] ,
\end{align}
where
\begin{gather}
tr[\gamma^5]=tr[\gamma^5\gamma^\mu\gamma^\nu]=0,\;\; tr[\gamma^5\gamma^\mu\gamma^\nu\gamma^\alpha\gamma^\beta]=-4\epsilon^{\mu\nu\alpha\beta}
\end{gather}
have been used.

Setting $\theta(x)=-c_\mu x^{\mu}$, we have
\begin{gather}
A_{\mu+}^{(s)}=A_\mu +c_\mu(1-s),\;\; A_{\mu-}^{(s)}=A_\mu-c_{\mu}(1-s) , \\
F_{+,\mu\nu}^{(s)}=F^{(s)}_{-,\mu\nu}=\partial_{\mu}A_\nu-\partial_\nu A_\mu=F_{\mu\nu} .
\end{gather}
Then, we obtain
\begin{gather}
\sum_n[\varphi_n^{(s)\dagger}(x)\gamma^5 \varphi_n^{(s)}(x)+\phi^{(s)\dagger}_n\gamma^5\phi_n^{(s)}(x)]=\frac{1}{16\pi^2}\epsilon^{\mu\nu\alpha\beta}F_{\mu\nu}F_{\alpha\beta} .
\end{gather}

Finally, the effective action changes as
\begin{align}
S_{WM}\rightarrow S_{WM}^{(s)}=\int d^4x\Big[\bar{\Psi}(x)i\slashed{D}^{(s)}\Psi(x)-\int_0^s ds c_\mu x^\mu \frac{i}{16\pi^2}\epsilon^{\mu\nu\alpha\beta}F_{\mu\nu}F_{\alpha\beta}\Big]
\end{align}
under the chiral transformation, where $\slashed{D}^{(s)}=\gamma^\mu(\partial_\mu+iA_\mu+ic_\mu(1-s)\gamma^5)$. Setting $s=1$, we obtain
\begin{gather}
S_{WM}^{(0)}=\int d^4x\Big[\bar{\Psi}(x)i\gamma^\mu(\partial_\mu +i A_\mu+ic_\mu \gamma^5)\Psi(x)\Big]\\
\Rightarrow S_{WM}^{(1)}=\int d^4x\Big[\bar{\Psi}(x)i\gamma^\mu(\partial_\mu +i A_\mu)\Psi(x)-\frac{i c_\mu x^\mu}{16\pi^2}\epsilon^{\mu\nu\alpha\beta}F_{\mu\nu}F_{\alpha \beta}\Big] .
\end{gather}

\section{Derivation of a Fermi arc surface state in the time-reversal symmetry-broken Weyl metal phase} \label{Appendix:surface-state-TRB Wm}

\subsection{Zero mass}

Following Ref. \cite{TRB WM Anomaly}, we derive a Fermi arc state. The reason why we show this procedure in this appendix is that this part is also applied to the case of inversion symmetry-broken Weyl metals.

We consider
\begin{align}
H_{WM}&=\int d^3 x \bar{\Psi}(x)[i\gamma^k\partial_k-c_\mu\gamma^\mu\gamma^5]\Psi(x)\nonumber\\
%
%
&=\int d^3 x \Psi^\dagger(x) \Bigg(\begin{array}{cc}-i\vec{\sigma}\cdot\nabla+\vec{\sigma}\cdot\mathbf{c} & 0 \\
0 & i\vec{\sigma}\cdot\nabla+\vec{\sigma}\cdot\mathbf{c} \end{array}\Bigg) \Psi(x) .
\end{align}
Setting $\mathbf{c}=c\theta(x)\hat{z}$, where the $x=0$ plane is a boundary, we have
\begin{align}
\Bigg(\begin{array}{cc} -i\vec{\sigma}\cdot\nabla+\sigma_3 c\theta(x) & 0 \\ 0 & i\vec{\sigma}\cdot\nabla+\sigma_3 c\theta(x) \end{array}\Bigg)\psi(x,y,z)=E\psi(x,y,z) .
\end{align}
Considering the translational symmetry in both directions of $y$ and $z$, we take $\psi(x,y,z)=e^{ik_y y +ik_z z} \phi_{k_y,k_z}(x)$. Then, we obtain
\begin{align}
&\Bigg(\begin{array}{cc}-i\sigma_1\partial_x+\sigma_2k_y+\sigma_3k_z+\sigma_3c\theta(x) & 0 \\
0 & i\sigma_1\partial_x-\sigma_2k_y-\sigma_3k_z+\sigma_3c\theta(x) \end{array}\Bigg)\phi_{k_y,k_z}(x)=E\phi_{k_y,k_z}(x) .
\end{align}

Inserting $\phi_{k_y,k_z}(x)=\Big(\begin{array}{c}u_{k_y,k_z}(x) \\ v_{k_y,k_z}(x)\end{array}\Big)$ into the above equation, we obtain
\begin{align}
&[-i\sigma_1\partial_x+\sigma_2k_y+\sigma_3k_z+\sigma_3c\theta(x)-E]u_{k_y,k_z}(x) = 0 , \\
&[i\sigma_1\partial_x-\sigma_2 k_y-\sigma_3 k_z+\sigma_3 c\theta(x)-E]v_{k_y,k_z}(x) = 0 .
\end{align}
Taking into account
\begin{gather}
u_{k_y,k_z}(x)=u^1_{k_y,k_z}(x)\Big(\begin{array}{c}1\\ 0\end{array}\Big)+u^2_{k_y,k_z}(x)\Big(\begin{array}{c}0\\ 1\end{array}\Big) , \\
v_{k_y,k_z}(x)=v^1_{k_y,k_z}(x)\Big(\begin{array}{c}1\\ 0\end{array}\Big)+v^2_{k_y,k_z}(x)\Big(\begin{array}{c}0\\ 1\end{array}\Big) ,
\end{gather}
we have
\begin{align}
&\Big\{\begin{array}{c} i(\partial_x-k_y)u^1+(E+k_z+c\theta(x))u^2=0\\
i(\partial_x+k_y)u^2+(E-k_z-c\theta(x))u^1=0\end{array} , \label{OeqU}\\
&\Big\{\begin{array}{c}i(\partial_x-k_y)v^1-(E-k_z+c\theta(x))v^2=0\\
i(\partial_x+k_y)v^2-(E+k_z-c\theta(x))v^1=0\end{array} . \label{OeqV}
\end{align}

%
%

Solving these equations of motion, we find
\begin{gather}
u^{i}(x)=\Bigg\{\begin{array}{cc} A^i e^{\sqrt{k_z^2+k_y^2-E^2}x}, & x<0 \\
A^i e^{-\sqrt{(k_z+c)^2+k_y^2-E^2}x}, & x>0\end{array},\;\; v^{i}(x)=\Bigg\{\begin{array}{cc} B^i e^{\sqrt{k_z^2+k_y^2-E^2}x}, & x<0 \\
B^i e^{-\sqrt{(k_z-c)^2+k_y^2-E^2}x}, & x>0\end{array} .
\end{gather}
Unknown coefficients are determined by the following boundary conditions
\begin{gather}
\lim_{\epsilon\rightarrow 0^+}u^{i}(\epsilon)=\lim_{\epsilon\rightarrow 0^{-}}u^{i}(\epsilon),\; lim_{\epsilon\rightarrow 0^+}v^{i}(\epsilon)=\lim_{\epsilon\rightarrow 0^{-}}v^{i}(\epsilon) , \label{Boundary_Condition1_TRSB_Zero_Mass} \\
\lim_{\epsilon\rightarrow 0^+}u'^{1}(\epsilon)-\lim_{\epsilon\rightarrow 0^-}u'^{1}(\epsilon)-icu^2(0)=0 , \label{Boundary_Condition2_TRSB_Zero_Mass} \\
\lim_{\epsilon\rightarrow 0^+}u'^{2}(\epsilon)-\lim_{\epsilon\rightarrow 0^-}u'^{2}(\epsilon)+icu^1(0)=0 , \label{Boundary_Condition3_TRSB_Zero_Mass} \\
\lim_{\epsilon\rightarrow 0^+}v'^{1}(\epsilon)-\lim_{\epsilon\rightarrow 0^-}v'^{1}(\epsilon)+icv^2(0)=0 , \label{Boundary_Condition4_TRSB_Zero_Mass} \\
\lim_{\epsilon\rightarrow 0^+}v'^{2}(\epsilon)-\lim_{\epsilon\rightarrow 0^-}v'^{2}(\epsilon)-icv^1(0)=0 , \label{Boundary_Condition5_TRSB_Zero_Mass}
\end{gather}
resulting in
\begin{gather}
\Big\{\begin{array}{c}A^1(\sqrt{(k_z+c)^2+k_y^2-E^2}+\sqrt{k_z^2+k_y^2-E^2})+icA^2=0\\
A^2(\sqrt{(k_z+c)^2+k_y^2-E^2}+\sqrt{k_z^2+k_y^2-E^2})-icA^1=0\end{array}\\
\Big\{\begin{array}{c}B^1(\sqrt{(k_z-c)^2+k_y^2-E^2}+\sqrt{k_z^2+k_y^2-E^2})-icB^2=0\\
B^2(\sqrt{(k_z-c)^2+k_y^2-E^2}+\sqrt{k_z^2+k_y^2-E^2})+icA^1=0\end{array} .
\end{gather}

%
%
%


A little algebra gives rise to
\begin{gather}
A^1=\sqrt{-\frac{k_z(c+k_z)}{c}},\;\; B^1=\sqrt{\frac{k_z(c-k_z)}{c}} ,
\end{gather}
where the normalization of $\int_{-\infty}^{\infty}(|u_{k_y,k_z}(x)|^2\text{ or }|v_{k_y,k_z}(x)|^2)dx=1$ has been used. As a result, we find a Fermi-arc state given by
\begin{enumerate}
\item[(i)] $-c<k_z<0$
\begin{gather}
\psi_{k_y,k_z}(x,y,z)=e^{ik_yy+ik_zz}\sqrt{\frac{-k_z(k_z+c)}{c}}\left(\begin{array}{c}1\\ i \\ 0\\ 0 \end{array}\right)[e^{-k_z x}(1-\theta(x))+e^{-(k_z+c)x}\theta(x)]\\
E=k_y
\end{gather}
\item[(ii)] $0<k_z<c$
\begin{gather}
\psi_{k_y,k_z}(x,y,z)=e^{ik_y y+ik_zz}\sqrt{\frac{k_z(c-k_z)}{c}}\left(\begin{array}{c}0\\ 0 \\ 1\\ -i\end{array}\right)[e^{k_zx}(1-\theta(x))+e^{-(c-k_z)x}\theta(x)]\\
E=k_y .
\end{gather}
\end{enumerate}

%
%

\subsection{Nonzero mass}

It is not difficult to generalize the above calculation in the presence of a mass term which preserves both time reversal and inversion symmetries. Since we do not find any concrete calculations in the presence of a mass term, we also show explicit procedures. We consider the following equation of motion for surface states
\begin{align}
H_{surf}\psi(x,y,z)=\Bigg(\begin{array}{cc} -i\vec{\sigma}\cdot\nabla+\sigma_3 c\theta(x) & m \\ m & i\vec{\sigma}\cdot\nabla+\sigma_3 c\theta(x) \end{array}\Bigg)\psi(x,y,z)=E\psi(x,y,z) ,
\end{align}
where $PH(\mathbf{k})P=H(-\mathbf{k})$ with $P=\tau_1\otimes 1$ can be checked easily. Introducing $\psi(x,y,z)=e^{ik_yy+ik_zz}\phi_{k_y,k_z}(x)$ into the above, we obtain
\begin{align}
&\Bigg(\begin{array}{cc}-i\sigma_1\partial_x+\sigma_2k_y+\sigma_3k_z+\sigma_3c\theta(x) & m \\ m & i\sigma_1\partial_x-\sigma_2k_y-\sigma_3k_z+\sigma_3c\theta(x) \end{array}\Bigg)\phi_{k_y,k_z}(x)=E\phi_{k_y,k_z}(x) .
\end{align}

Following the previous section, we obtain
\begin{align}
&i(\partial_x-k_y)u^1+(E+k_z+c\theta(x))u^2-mv^2=0\label{FiniteMassEq1}\\
&i(\partial_x+k_y)u^2+(E-k_z-c\theta(x))u^1-mv^1=0\\
&i(\partial_x-k_y)v^1-(E-k_z+c\theta(x))v^2+mu^2=0\\
&i(\partial_x+k_y)v^2-(E+k_z-c\theta(x))v^1+mu^1=0 . \label{FiniteMassEq4}
\end{align}
Introducing $D\equiv i(\partial_x-k_y)$ and $D^\dagger\equiv i(\partial_x+k_y)$ into the above equations and performing a little algebra, we obtain
%
%
\begin{gather}
[D^\dagger D -E^2+(k_z+c\theta(x))^2+m^2]u^1+2mc\theta(x)v^1+ic\delta(x)u^2=0\\
[D^\dagger D -E^2+(k_z-c\theta(x))^2+m^2]v^1+2mc\theta(x)u^1-ic\delta(x)v^2=0\\
[DD^\dagger-E^2+(k_z+c\theta(x))^2+m^2]u^2-2mc\theta(x)v^2-ic\delta(x)u^1=0\\
[DD^\dagger-E^2+(k_z-c\theta(x))^2+m^2]v^2-2mc\theta(x)u^2+ic\delta(x)v^1=0 .
\end{gather}
Boundary conditions are the same as those of the zero-mass case. As a result, we find
\begin{enumerate}
\item[(i)] $x<0$
\begin{gather}
[\partial_x^2+E^2-k_y^2-k_z^2-m^2]u^{1/2}/ v^{1/2}=0\\
u^{1/2}=A^{1/2}e^{\sqrt{k_z^2+k_y^2+m^2-E^2}x},\;\; v^{1/2}=\tilde{A}^{1/2}e^{\sqrt{k_z^2+k_y^2+m^2-E^2}x}\\
E^2<k_z^2+k_y^2+m^2
\end{gather}
\item[(ii)] $x>0$
\begin{align}
&[\partial^2+E^2-k_y^2-k_z^2-m^2-c^2-2c\sqrt{k_z^2+m^2}]\nonumber\\
&\times[\partial_x^2+E^2-k_y^2-k_z^2-m^2-c^2+2c\sqrt{k_z^2+m^2}]u^{1/2}/v^{1/2}=0\\
&u^{1/2}=B^{1/2}e^{-\sqrt{(\sqrt{k_z^2+m^2}+c)^2+k_y^2-E^2}x}+C^{1/2}e^{-\sqrt{(\sqrt{k_z^2+m^2}-c)^2+k_y^2-E^2}x},\\
&v^{1/2}=\tilde{B}^{1/2}e^{-\sqrt{(\sqrt{k_z^2+m^2}+c)^2+k_y^2-E^2}x}+\tilde{C}^{1/2}e^{-\sqrt{(\sqrt{k_z^2+m^2}-c)^2+k_y^2-E^2}x} .
\end{align}
\end{enumerate}

Applying the boundary conditions Eqs. (\ref{Boundary_Condition1_TRSB_Zero_Mass}) $\sim$ (\ref{Boundary_Condition5_TRSB_Zero_Mass}) into the above formal solutions, we obtain
\begin{gather}
A^{1/2}=B^{1/2}+C^{1/2},\;\; \tilde{A}^{1/2}=\tilde{B}^{1/2}+\tilde{C}^{1/2}\\
B^1\sqrt{(K_z+c)^2+k_y^2-E^2}+C^1\sqrt{(K_z-c)^2+k_y^2-E^2}+A^1\sqrt{k_y^2+k_z^2+m^2-E^2}+icA^2=0\label{FiniteMassBC1}\\
B^2\sqrt{(K_z+c)^2+k_y^2-E^2}+C^2\sqrt{(K_z-c)^2+k_y^2-E^2}+A^2\sqrt{k_y^2+k_z^2+m^2-E^2}-icA^1=0\\
\tilde{B}^1\sqrt{(K_z+c)^2+k_y^2-E^2}+\tilde{C}^1\sqrt{(K_z-c)^2+k_y^2-E^2}+\tilde{A}^1\sqrt{k_y^2+k_z^2+m^2-E^2}-ic\tilde{A}^2=0\\
\tilde{B}^2\sqrt{(K_z+c)^2+k_y^2-E^2}+\tilde{C}^2\sqrt{(K_z-c)^2+k_y^2-E^2}+\tilde{A}^2\sqrt{k_y^2+k_z^2+m^2-E^2}+ic\tilde{A}^1=0 , \label{FiniteMassBC4}
\end{gather}
where $K_z=\sqrt{k_z^2+m^2}$.
Although it is a little bit complex to solve these equations, we find
\begin{gather}
B^+=B^-=C^+=\tilde{B}^+=\tilde{B}^-=\tilde{C}^-=0\\
A^+=\tilde{A}^{-}=0,\;\; A^-=C^-,\;\; \tilde{A}^+=\tilde{C}^+\\
\Rightarrow B^1=B^2=\tilde{B}^1=\tilde{B}^2=0\\
A^2=iA^1,\;\; \tilde{A}^2=-i\tilde{A}^1,\;\; C^2=iC^1,\;\; \tilde{C}^2=-i\tilde{C}^1,\;\; A^1=C^1,\;\; \tilde{A}^{1}=\tilde{C}^1\\
\tilde{A}^1=-\frac{m}{\sqrt{k_z^2+m^2}-k_z}A^1\\
\Rightarrow \left\{\begin{array}{cc}A^2=iA^1, & \tilde{A}^2=-i\tilde{A}^1 \\ C^1=A^1, & \tilde{C}^1=\tilde{A}^1 \\ C^2=iA^1, & \tilde{C}^2=-i\tilde{A}^1\end{array}\right\},\; \tilde{A}^1=-\frac{m}{\sqrt{k_z^2+m^2}-k_z}A^1.
\end{gather}
Finally, we obtain a Fermi-arc state given by
\begin{enumerate}
\item[(i)] $x<0$
\begin{gather}
-\sqrt{c^2-m^2}<k_z<\sqrt{c^2-m^2}\\
 \Big\{\begin{array}{cc}u^1=A^1e^{\sqrt{k_z^2+m^2}x}, & u^2=iA^1e^{\sqrt{k_z^2+m^2}x} \\
v^1=-\frac{m}{\sqrt{k_z^2+m^2}-k_z}A^1e^{\sqrt{k_z^2+m^2}x}, & v^2=i\frac{m}{\sqrt{k_z^2+m^2}-k_z}A^1e^{\sqrt{k_z^2+m^2}x}
\end{array}
\end{gather}
\item[(ii)] $x>0$
\begin{gather}
-\sqrt{c^2-m^2}<k_z<\sqrt{c^2-m^2}\\
\Big\{\begin{array}{cc}u^1=A^1e^{-(c-\sqrt{k_z^2+m^2})x}, & u^2=iA^1e^{-(c-\sqrt{k_z^2+m^2})x} \\
v^1=-\frac{m}{\sqrt{k_z^2+m^2}-k_z}A^1e^{-(c-\sqrt{k_z^2+m^2})x}, & v^2=i\frac{m}{\sqrt{k_z^2+m^2}-k_z}A^1e^{-(c-\sqrt{k_z^2+m^2})x} , \end{array}
\end{gather}
\end{enumerate}
where the energy eigen value is $E=k_y$.

In summary, the Fermi-arc state is given by
\begin{gather}
-\sqrt{c^2-m^2}<k_z<\sqrt{c^2-m^2},\; E=k_y\\
 \psi(x,y,z)_{k_y,k_z}=A^1e^{ik_y y+ik_z z}\left(\begin{array}{c}1 \\ i \\ -\frac{m}{\sqrt{k_z^2+m^2}-k_z}\\ i\frac{m}{\sqrt{k_z^2+m^2}-k_z}\end{array}\right)e^{(-c\theta(x)+\sqrt{k_z^2+m^2})x} ,
\end{gather}
where $A^1=\Big(\frac{(\sqrt{k_z^2+m^2}-k_z)^2\sqrt{k_z^2+m^2}(c-\sqrt{k_z^2+m^2})}{2c(k_z^2+m^2-k_z\sqrt{k_z^2+m^2})}\Big)^{1/2}$ determined by the normalization condition.

\subsection{Chirality of the surface state}

One can define the chirality operator for a surface state: $\bar{\gamma}=\gamma^0\gamma^2=\left(\begin{array}{cc}-\sigma_2 & 0 \\ 0 & \sigma_2\end{array}\right)$. Since boundary excitations are propagating only along the $y$-direction, $\sigma_2$ appears. Application of the chirality operator to the surface state gives rise to
\begin{gather}
\bar{\gamma}\phi_{k_y,k_z}=-\phi_{k_y,k_z} .
\end{gather}
Boundary excitations have definite chirality.

\section{Gauge anomaly of the U(1) surface current in time-reversal symmetry-broken Weyl metals} \label{Appendix:1-loop-correction-TRB WM}

We introduce a bosonic ``spinor" $\phi(x)$ into an effective action of the surface Fermi-arc state as follows
\begin{gather}
S=\int d^2x \Big[\bar{\Psi}(x)i\gamma^\mu(\partial_\mu+ieA_\mu \mathcal{P}_-)\Psi(x)+\bar{\phi}(x)i\gamma^\mu(\partial_\mu+ieA_\mu\mathcal{P}_-)\phi(x)+\bar{\phi}(x)M\phi(x)\Big] ,
\end{gather}
where $M$ is the mass of the bosonic spinor field. Recall that $\mathcal{P}_-$ is the chirality projection operator. This is referred to as the Pauli-Villars regularization \cite{Anomalies in QFT}. If we consider the chiral gauge transformation for $\bar{\phi}(x)$ and $\phi(x)$, they transform as $\bar{\Psi}(x)$ and $\Psi(x)$ and the mass term breaks the chiral gauge symmetry explicitly.

Performing the Fourier transformation, we obtain
\begin{align}
S=\int \frac{d^2k}{(2\pi)^2}\Big[\bar{\Psi}(k)\slashed{k}\Psi(k)+\bar{\phi}(k)(\slashed{k}+M)\phi(k) - \int \frac{d^2q}{(2\pi)^2}\Big(\bar{\Psi}(k+q)\slashed{A}(q)\mathcal{P}_-\Psi(k)+\bar{\phi}(k+q)\slashed{A}(q)\mathcal{P}_-\phi(k)\Big)\Big] ,
\end{align}
where
\begin{gather}
\Psi(x)=\int \frac{d^2k}{(2\pi)^2}e^{-ik\cdot x}\Psi(x),\;\; \bar{\Psi}(x)=\int \frac{d^2k}{(2\pi)^2}e^{ik \cdot x}\bar{\Psi}(k),\\
\phi(x)=\int \frac{d^2k}{(2\pi)^2}e^{-ik\cdot x}\phi(x),\;\; \bar{\phi}(x)=\int \frac{d^2k}{(2\pi)^2}e^{ik \cdot x}\bar{\phi}(k),\\
A_\mu(q)=\int d^2x e^{iq\cdot x}A_\mu(x) .
\end{gather}
Accordingly, Green functions are given by
\begin{gather}
G(k)=\langle \Psi(k)\bar{\Psi}(k)\rangle=\frac{\slashed{k}}{k^2},\;\; \tilde{G}(k)= \langle \phi(k)\bar{\phi}(k)\rangle=\frac{\slashed{k}-M}{k^2-M^2} .
\end{gather}

The current operator is
\begin{gather}
j^\mu(q)=\int d^2x e^{-iq\cdot x}j^\mu(x)=\int d^2xe^{-iq\cdot x}\bar{\Psi}(x)\gamma^\mu\mathcal{P}_-\Psi(x)=\int \frac{d^2k}{(2\pi)^2}\bar{\Psi}(k+q)\gamma^\mu\mathcal{P}_-\Psi(k)
\end{gather}
under the Fourier transformation. Applying the Pauli-Villars regularization into the above expression, we obtain
\begin{gather}
j_{reg}^\mu(q)=\int \frac{d^2k}{(2\pi)^2}\Big[\bar{\Psi}(k+q)\gamma^\mu \mathcal{P}_- \Psi(k)+\bar{\phi}(k+q)\gamma^\mu \mathcal{P}_-\phi(k)\Big] .
\end{gather}

Up to the one-loop order, we find
\begin{align}
\langle j_{reg}^\mu(q)\rangle&=\lim_{M^2\rightarrow \infty}\int \frac{d^2k}{(2\pi)^2}\Big[-tr(G(k)\gamma^\mu\mathcal{P}_-G(k+q)\gamma^\nu\mathcal{P}_-)+tr(\tilde{G}(k)\gamma^\mu\mathcal{P}_-\tilde{G}(k+q)\gamma^\nu \mathcal{P}_-)\Big]A_\nu(-q)\nonumber\\
&= -\Big[2(2q^\mu q^\nu-g^{\mu\nu }q^2)+i\epsilon^{\mu\nu}(q_\mu^2-q_\nu^2)+2i\sum_{\alpha}\delta_{\mu\nu}q_\alpha\epsilon^{\alpha\nu}q_\mu\Big]\frac{A_\nu(-q)}{4\pi q^2} .
\end{align}
The difference in the sign comes from the fact whether the particle is a fermion or a boson. Here, we have used the following properties of $\gamma^\mu$ and the integral identity:
\begin{gather}
tr(\gamma^\mu)=tr(\bar{\gamma})=0,\;\; tr(\gamma^\mu\gamma^\nu)=4g^{\mu\nu}\\
tr(\gamma^\mu\gamma^\nu\gamma^\alpha)=0,\;\; tr(\gamma^\mu\gamma^\nu\bar{\gamma})=\epsilon^{\mu\nu}2i\; (\epsilon^{zt}=1)\\
tr(\gamma^\alpha\gamma^\beta\gamma^\mu\gamma^\nu)=4(g^{\alpha\beta}g^{\mu\nu}-g^{\alpha\mu}g^{\beta\nu}+g^{\alpha\nu}g^{\beta\mu})\\
tr(\gamma^\mu\bar{\gamma})=tr(\gamma^\alpha\gamma^\beta\gamma^\mu\bar{\gamma})=0\\
tr(\gamma^\alpha\gamma^\beta\gamma^\mu\gamma^\nu\bar{\gamma})=-2i[\epsilon^{\mu\nu}\delta_{\alpha\beta}(\delta_{\beta\mu}+\delta_{\beta\nu})+\epsilon^{\alpha\beta}\delta_{\mu\nu}(\delta_{\alpha\mu}+\delta_{\beta\mu})]\\
\int \frac{d^2k}{(2\pi)^2}\frac{1}{(k^2-\Delta)^n}=\frac{1}{4\pi}\frac{1}{n-1}\frac{1}{\Delta^{n-1}} \;\;(n>1) .
\end{gather}
As a result, we find that the surface U(1) current given by the Fermi-arc state is not conserved
\begin{gather}
q\cdot \langle j_{reg}(q)\rangle= \frac{i \epsilon^{\mu\nu}q_\mu A_\nu(-q)}{4\pi}\Rightarrow \partial_\mu j^{\mu}(x) =\frac{i}{8\pi}\epsilon^{\mu\nu}F_{\mu\nu} .
\end{gather}

\section{The total Berry flux density should vanish in the time reversal symmetric system} \label{Appendix:Zero-Berry-flux-IB WM}

Suppose a two-band system which consists of two eigenstates; $|n_{I}(\mathbf{k})\rangle$ and $|n_{II}(\mathbf{k})\rangle$. These eigenstates are related with time reversal symmetry; $\mathcal{T}|n_{I}(\mathbf{k})\rangle=|n_{II}(-\mathbf{k})\rangle$ and $\mathcal{T}|\mathbf{n}_{II}(\mathbf{k})\rangle=(\mathcal{T})^2 |\mathbf{n}_{I}(-\mathbf{k})\rangle$.

Taking into account the Berry connection for each band and the following consideration
\begin{align}
\mathcal{A}_{I}(\mathbf{k})&=i\langle n_{I}(\mathbf{k})|\nabla_k|n_{I}(\mathbf{k})\rangle \\
\mathcal{A}_{II}(\mathbf{k})&=i\langle n_{II}(\mathbf{k})|\nabla_k|n_{II}(\mathbf{k})\rangle= i\langle \mathcal{T}n_{I}(-\mathbf{k})|\nabla_k|\mathcal{T}n_{I}(-\mathbf{k})\rangle\nonumber \\
&=i\langle \mathcal{T}n_{I}(-\mathbf{k})|\mathcal{T}\nabla_kn_{I}(-\mathbf{k})\rangle \;(\because [\nabla_k,\mathcal{T}]=0)\nonumber \\
&=i\langle \nabla_kn_I(-\mathbf{k})|n_{I}(-\mathbf{k})\rangle\;(\because \langle \mathcal{T}a|\mathcal{T}b\rangle=\langle b|a\rangle)\nonumber \\
&=-i\langle n_{I}(-\mathbf{k})|\nabla_k|n_I(-\mathbf{k})\rangle=i\langle n_{I}(-\mathbf{k})|\nabla_{-k}|n_{i}(-\mathbf{k})=\mathcal{A}_{I}(-\mathbf{k})\\
\mathcal{B}_{I}(\mathbf{k})&=\nabla_k\times \mathcal{A}_I(\mathbf{k})\\
\mathcal{B}_{II}(\mathbf{k})&=\nabla_k\times\mathcal{A}_{II}(\mathbf{k})=-\nabla_{-k}\times \mathcal{A}_{I}(-\mathbf{k})=-\mathcal{B}_{I}(-\mathbf{k}) ,
\end{align}
we confirm $\int_{S}(\mathcal{B}_I+\mathcal{B}_{II})\cdot dS=0$.

\section{Inversion and Time-reversal transformation operators for the low-energy effective Hamiltonian} \label{Appendix:Inversion-TimeReversal-Operator}

We point out that the inversion transformation operator is represented in the original-basis state $|\mathbf{k},\sigma,s\rangle$ as follows
\begin{align}
\mathcal{P}\mathcal{H}\mathcal{P}^{-1}&=\sum_{\mathbf{k}}\sum_{i,j}\mathcal{P}|\mathbf{k},i\rangle H_{ij}(\mathbf{k})\langle \mathbf{k},j|\mathcal{P}^{-1} =\sum_{\mathbf{k}}\sum_{i,j}|-\mathbf{k},i\rangle H_{ij}(-\mathbf{k})\langle -\mathbf{k},j|=\mathcal{H} \nonumber \\
\Rightarrow &P_{ij}=\langle -\mathbf{k},i|\mathbf{k},j\rangle,\;\; P_{i\alpha}H_{\alpha\beta}(\mathbf{k})P^{-1}_{\beta j}=H_{ij}(-\mathbf{k})\\
\Rightarrow &P=\sigma_z : \left(\begin{array}{c}\mathcal{P}|\mathbf{k},\sigma=+,s\rangle=|-\mathbf{k},\sigma=+,s\rangle \\ \mathcal{P}|\mathbf{k},\sigma=-,s\rangle=-|-\mathbf{k},\sigma=-,s\rangle \end{array}
\right) .
\end{align}
Note that the operator $\mathcal{P}$ relates $|\mathbf{k},i\rangle$ with $|-\mathbf{k},i\rangle$. Here, we find the representation for $\mathcal{P}$ in the enlarged Hilbert space $|\mathbf{k},\sigma,s,\tau\rangle$.

Considering
\begin{gather}
\mathcal{P}|\mathbf{k},\sigma,s\rangle\approx \mathcal{P}|\mathbf{k},\sigma,s,+\rangle \propto |-\mathbf{k},\sigma,s\rangle\approx |-\mathbf{k},\sigma,s,-\rangle ,
\end{gather}
we obtain the representation of $\mathcal{P}$ in the enlarged Hilbert space, given by
\begin{gather}
\mathcal{P}\left(\begin{array}{c}|\mathbf{k},\sigma=+,s,\tau=+\rangle \\ |\mathbf{k},\sigma=-,s,\tau=+\rangle \\|\mathbf{k},\sigma=+,s,\tau=-\rangle \\ |\mathbf{k},\sigma=-,s,\tau=-\rangle  \end{array}\right)=\left(\begin{array}{c}|-\mathbf{k},\sigma=+,s,\tau=-\rangle \\ -|-\mathbf{k},\sigma=-,s,\tau=-\rangle \\|-\mathbf{k},\sigma=+,s,\tau=+\rangle \\ -|-\mathbf{k},\sigma=-,s,\tau=+\rangle  \end{array}\right)\Rightarrow \tilde{P}=\sigma^z\otimes \tau^x .
\end{gather}
In the same way we find the representation for the time-reversal transformation as well
\begin{gather}
\tilde{T}=is^y\otimes \tau^x \mathcal{K}.
\end{gather}

\section{Derivation of an effective axionic action for inversion symmetry-broken Weyl metals} \label{Appendix:Axionic-Action-IB-Weyl}

\subsection{$[\tau^z,\Gamma^\mu]=0$ and $\{s^z,\Gamma^\mu\}=0$}

The procedure is quite similar to the case of time-reversal symmetry-broken Weyl metals. Since any explicit calculations have not been shown for inversion symmetry-broken Weyl metals as far as we know, we report all detailed steps in this appendix. Taking into account the chiral gauge transformation in this case [Eq. (\ref{GammaAction})], we obtain
\begin{gather}
\Psi\rightarrow e^{i\Gamma^5\tau^z\beta(x)},\; \bar{\Psi}\rightarrow \bar{\Psi}e^{i\Gamma^5\tau^z\beta(x)}\;(\because [\tau^z,\Gamma^\mu]=0)\\
S\rightarrow S-\int d^4x\bar{\Psi}(x)\Gamma^\mu\Gamma^5\tau^z\partial_\mu \beta(x)\Psi(x)\\
\Rightarrow S'=\int d^4x \bar{\Psi}(x)[i\Gamma^\mu\partial_\mu-(\alpha+\partial_1\beta)\Gamma^1\Gamma^5\tau^z+k_0\Gamma^3\Gamma^5s^z]\Psi(x) ,
\end{gather}
where we set $\partial_\mu \beta=\delta^{\mu1}\partial_1\beta$. Considering $\beta(x)=ds\theta_v(x)$ and performing multiple steps of chiral rotations as discussed in the previous section, we obtain
\begin{gather}
S^{(s)}=\int d^4x\bar{\Psi}(x)[i\Gamma^\mu\partial_\mu-(\alpha+s\partial_1\theta_v)\Gamma^1\Gamma^5\tau^z+k_0\Gamma^3\Gamma^5s^z]\Psi(x)\equiv \int d^4x\bar{\Psi}(x)i\slashed{D}^{(s)}\Psi(x)\\
\slashed{D}^{(s)}=\Gamma^\mu\partial_\mu+i(\alpha+s\partial_1\theta_v)\Gamma^1\Gamma^5\tau^z-ik_0\Gamma^3\Gamma^5s^z .
\end{gather}

In order to find an effective action involved with the $\alpha$ term, we introduce two types of gauge fields in addition to the conventional U(1) gauge field $A_{\mu}$: the spin gauge field $S_{\mu}$ and the valley gauge field $V_{\mu}$ as follows
\begin{gather}
\slashed{D}^{(s)}=\Gamma^\mu(\partial_\mu+iA_\mu+is^zS_\mu+i\tau^zV_{\mu})+i(\alpha+s\partial_1\theta_v)\Gamma^1\Gamma^5\tau^z-ik_0\Gamma^3\Gamma^5s^z .
\end{gather}

Following the previous section, we calculate the change of the integral measure under this chiral rotation and obtain an effective action
\begin{align}
S_{eff}^{(s)}=S^{(s)}+\int d^4x\int_0^sds \theta_v(x)i\sum_n(\varphi_n^{(s)\dagger}\Gamma^5\tau^z\varphi_n^{(s)}+\phi_n^{(s)\dagger}\Gamma^5\tau^z\phi_n^{(s)})
\end{align}
where
\begin{gather}
\slashed{D}^{(s)\dagger}\slashed{D}^{(s)}\varphi_n^{(s)}(x)=\lambda_n^2 \varphi_n^{(s)}(x),\;\;\slashed{D}^{(s)}\slashed{D}^{(s)\dagger}\phi_n^{(s)}(x)=\lambda_n^2 \phi_n^{(s)}(x) \\
\slashed{D}^{(s)}\varphi_n^{(s)}(x)=\lambda_n\phi_n^{(s)}(x),\;\; \slashed{D}^{(s)\dagger}\phi_n^{(s)}(x)=\lambda_n \varphi_n^{(s)}(x) .
\end{gather}
The change of the integral measure can be evaluated in the following way
\begin{align}
&\sum_n[\varphi_n^{(s)\dagger}(x)\gamma_v^5\varphi_n^{(s)}(x)+\phi^{(s)\dagger}_n\gamma_v^5\phi_n^{(s)}(x)]=\lim_{M\rightarrow \infty}\int \frac{d^4 k}{(2\pi)^4}e^{-ik\cdot x}tr\Big[\gamma_v^5\Big( e^{-\frac{(\slashed{D}_+^{(s)})^2}{M^2}}+e^{-\frac{(\slashed{D}^{(s)}_-)^2}{M^2}}\Big)\Big]e^{ik\cdot x}\nonumber\\
&=\lim_{M\rightarrow \infty}\int\frac{d^4k}{(2\pi)^4}e^{-\frac{k_\mu^2}{M^2}}tr\Big[\gamma_v^5\Big(e^{-\frac{-(D_{+\mu}^{(s)})^2-2ik_\mu D_{+\mu}^{(s)}+\frac{i}{4}[\Gamma^\mu,\Gamma^\nu]F_{+,\mu\nu}^{(s)}}{M^2}}+e^{-\frac{-(D_{-\mu}^{(s)})^2-2ik_\mu D_{-\mu}^{(s)}+\frac{i}{4}[\Gamma^\mu,\Gamma^\nu]F_{-,\mu\nu}^{(s)}}{M^2}}\Big)\Big]\nonumber\\
&=\lim_{M\rightarrow\infty}\int\frac{d^4k}{(2\pi)^4}e^{-k_\mu^2}tr\Big[-\frac{1}{16}\gamma_v^5\Big([\Gamma^\mu,\Gamma^\nu](F_{\mu\nu}+\tau^zF_{v,\mu\nu})\Big)^2\Big]\nonumber \\
&=-\frac{1}{4}\lim_{M\rightarrow\infty}\int\frac{d^4k}{(2\pi)^4}e^{-k_\mu^2}tr\Big[\gamma_v^5\Gamma^\mu\Gamma^\nu\Gamma^\alpha\Gamma^\beta\tau^z (F_{v,\mu\nu}F_{\mu\nu}+F_{\mu\nu}F_{v,\alpha\beta})\Big]=\frac{1}{4\pi^2}\epsilon^{\mu\nu\alpha\beta}F_{v,\mu\nu}F_{\alpha\beta} ,
\end{align}
where
\begin{gather}
\slashed{D}^{(s)}=\slashed{D}_{+}^{(s)}P_{v,+}+\slashed{D}_{-}^{(s)}P_{v,-}\\
\slashed{D}_{\pm}^{(s)}=\Gamma^\mu(\partial_\mu+iA_{\pm,\mu}^{(s)})\\
A_{\pm,\mu}^{(s)}=A_\mu+\tau^zV_\mu+k_0\delta^{\mu3}\tau^z\pm[-S_\mu+\delta^{\mu1}(\alpha+s\partial_1\theta_v)]\\
F^{(s)}_{\pm,\mu\nu}=\partial_\mu A^{(s)}_{\pm,\nu}-\partial_\nu A^{(s)}_{\pm,\mu}=F_{\mu\nu}+\tau^zF_{v,\mu\nu}\mp F_{s,\mu\nu}\\
P_{v,\pm}=\frac{1\pm\gamma_v^5}{2} .
\end{gather}
$F_{\mu\nu}$ is the field strength tensor of the U(1) gauge field and $F_{v,\mu\nu}$ is that of the valley gauge field. Here, we have used the representation of $\Gamma^\mu$ matrices in terms of $\gamma^\mu$ and $\tau^\mu$ matrices ($\gamma_v^4=-i\gamma_v^0$) since it is more convenient for calculations. We also used
\begin{gather}
tr[\gamma_v^5]=tr[\gamma_v^5\gamma_v^\mu\gamma_v^\nu]=0\\ tr[\gamma_v^5\gamma_v^\mu\gamma_v^\nu\gamma_v^\alpha\gamma_v^\beta]=-4\epsilon^{\mu\nu\alpha\beta}\\
tr[A\otimes B]=tr[A]tr[B].
\end{gather}

Finally, we reach the following expression after the chiral transformation
\begin{align}
\therefore S_{eff}^{(s)}&=\int d^4x\bar{\Psi}(x)[i\Gamma^\mu\partial_\mu-(\alpha+s\partial_1\theta_v)\Gamma^1\Gamma^5\tau^z+k_0\Gamma^3\Gamma^5s^z]\Psi(x)\nonumber \\
&+i\int d^4x\int_0^sds \frac{\theta_v}{4\pi^2}\epsilon^{\mu\nu\alpha\beta}F_{v,\mu\nu}F_{\alpha\beta} .
\end{align}
If we set $\theta_v=-\alpha x^1$ with $s=1$, we obtain
\begin{gather}
S_{eff}^{(1)}=\int d^4x\bar{\Psi}(x)[i\Gamma^\mu\partial_\mu+k_0\Gamma^3\Gamma^5s^z]\Psi(x)-i \int d^4x\frac{\alpha x^1}{4\pi^2}\epsilon^{\mu\nu\alpha\beta}F_{v,\mu\nu}F_{\alpha\beta} .
\end{gather}
In other words, we find
\begin{gather}
S_{eff}^{v}\equiv -\int d^4x \frac{i\alpha x^1}{4\pi^2}\epsilon^{\mu\nu\alpha\beta}F_{v,\mu\nu}F_{\alpha\beta} ,
\end{gather}
where both charge and valley gauge fields are involved. Note that the coefficient of the effective action is four times larger than that of the time-reversal symmetry-broken Weyl metal. This point turns out to be essential in the anomaly cancelation.

\subsection{$\{\tau^z,\Gamma^\mu\}=0$ and $[s^z,\Gamma^\mu]= 0$}

Now, we consider the other chiral rotation and obtain
\begin{gather}
\Psi\rightarrow e^{i\Gamma^5s^z\beta(x)},\; \bar{\Psi}\rightarrow \bar{\Psi}e^{i\Gamma^5s^z\beta(x)}\;(\because [\tau^z,\Gamma^\mu]=0)\\
S\rightarrow S-\int d^4x\bar{\Psi}(x)\Gamma^\mu\Gamma^5s^z\partial_\mu \beta(x)\Psi(x)\\
\Rightarrow S'=\int d^4x \bar{\Psi}(x)[i\Gamma^\mu\partial_\mu-\alpha\Gamma^1\Gamma^5\tau^z+(k_0-\partial_3\beta)\Gamma^3\Gamma^5s^z]\Psi(x) ,
\end{gather}
where we set $\partial_\mu \beta=\delta^{\mu3}\partial_3\beta$. Taking $\beta(x)=ds\theta_s(x)$ and performing essentially the same steps of chiral rotations before, we obtain
\begin{gather}
S^{(s)}=\int d^4x\bar{\Psi}(x)[i\Gamma^\mu\partial_\mu-\alpha\Gamma^1\Gamma^5\tau^z+(k_0-s\partial_3\theta_s)\Gamma^3\Gamma^5s^z]\Psi(x)\equiv \int d^4x\bar{\Psi}(x)i\slashed{D}^{(s)}\Psi(x)\\
\slashed{D}^{(s)}=\Gamma^\mu\partial_\mu+i\alpha\Gamma^1\Gamma^5\tau^z-i(k_0-s\partial_3\theta_v)\Gamma^3\Gamma^5s^z .
\end{gather}
We also introduce the whole set of U(1) gauge fields, given by
\begin{gather}
\slashed{D}^{(s)}=\Gamma^\mu(\partial_\mu+iA_\mu+is^zS_\mu+i\tau^zV_\mu)+i\alpha\Gamma^1\Gamma^5\tau^z-i(k_0-s\partial_3\theta_s)\Gamma^3\Gamma^5s^z .
\end{gather}

Considering the change of the integral measure under this chiral rotation, we find the following effective action
\begin{align}
S_{eff}^{(s)}=S^{(s)}+\int d^4x\int_0^sds \theta_v(x)i\sum_n(\varphi_n^{(s)\dagger}\Gamma^5s^z\varphi_n^{(s)}+\phi_n^{(s)\dagger}\Gamma^5s^z\phi_n^{(s)}) ,
\end{align}
where
\begin{gather}
\slashed{D}^{(s)\dagger}\slashed{D}^{(s)}\varphi_n^{(s)}(x)=\lambda_n^2 \varphi_n^{(s)}(x),\;\;\slashed{D}^{(s)}\slashed{D}^{(s)\dagger}\phi_n^{(s)}(x)=\lambda_n^2 \phi_n^{(s)}(x) \\
\slashed{D}^{(s)}\varphi_n^{(s)}(x)=\lambda_n\phi_n^{(s)}(x),\;\; \slashed{D}^{(s)\dagger}\phi_n^{(s)}(x)=\lambda_n \varphi_n^{(s)}(x) .
\end{gather}
It is essentially the same procedure to evaluate the change of the integral measure as follows
\begin{align}
&\sum_n[\varphi_n^{(s)\dagger}(x)\gamma_s^5\varphi_n^{(s)}(x)+\phi^{(s)\dagger}_n\gamma_s^5\phi_n^{(s)}(x)]=\lim_{M\rightarrow \infty}\int \frac{d^4 k}{(2\pi)^4}e^{-ik\cdot x}tr\Big[\gamma_s^5\Big( e^{-\frac{(\slashed{D}_+^{(s)})^2}{M^2}}+e^{-\frac{(\slashed{D}^{(s)}_-)^2}{M^2}}\Big)\Big]e^{ik\cdot x}\nonumber\\
&=\lim_{M\rightarrow \infty}\int\frac{d^4k}{(2\pi)^4}e^{-\frac{k_\mu^2}{M^2}}tr\Big[\gamma_s^5\Big(e^{-\frac{-(D_{+\mu}^{(s)})^2-2ik_\mu D_{+\mu}^{(s)}+\frac{i}{4}[\Gamma^\mu,\Gamma^\nu]F_{+,\mu\nu}^{(s)}}{M^2}}+e^{-\frac{-(D_{-\mu}^{(s)})^2-2ik_\mu D_{-\mu}^{(s)}+\frac{i}{4}[\Gamma^\mu,\Gamma^\nu]F_{-,\mu\nu}^{(s)}}{M^2}}\Big)\Big]\nonumber\\
&=\lim_{M\rightarrow\infty}\int\frac{d^4k}{(2\pi)^4}e^{-k_\mu^2}tr\Big[-\frac{1}{16}\gamma_s^5\Big([\Gamma^\mu,\Gamma^\nu](F_{\mu\nu}+s^zF_{s,\mu\nu})\Big)^2\Big]\nonumber \\
&=-\frac{1}{4}\lim_{M\rightarrow\infty}\int\frac{d^4k}{(2\pi)^4}e^{-k_\mu^2}tr\Big[\gamma_s^5\Gamma^\mu\Gamma^\nu\Gamma^\alpha\Gamma^\beta s^z(F_{s,\mu\nu}F_{\mu\nu}+F_{\mu\nu}F_{s,\alpha\beta})\Big]=\frac{1}{4\pi^2}\epsilon^{\mu\nu\alpha\beta}F_{s,\mu\nu}F_{\alpha\beta} ,
\end{align}
where
\begin{gather}
\slashed{D}^{(s)}=\slashed{D}_{+}^{(s)}P_{s,+}+\slashed{D}_{-}^{(s)}P_{s,-}\\
\slashed{D}_{\pm}^{(s)}=\Gamma^\mu(\partial_\mu+iA_{\pm,\mu}^{(s)})\\
A_{\pm,\mu}^{(s)}=A_\mu+s^z(S_\mu-\alpha\delta^{\mu1})\mp[V_\mu+(k_0-s\partial_3\theta_v)\delta^{\mu 3}]\\
F^{(s)}_{\pm,\mu\nu}=\partial_\mu A^{(s)}_{\pm,\nu}-\partial_\nu A^{(s)}_{\pm,\mu}=F_{\mu\nu}+s^zF_{s,\mu\nu}\mp F_{v,\mu\nu}\\
P_{s,\pm}=\frac{1\pm\gamma_s^5}{2}.
\end{gather}
Here, $F_{s,\mu\nu}$ is the field strength tensor of the spin gauge field.

As a result, we find
\begin{align}
S_{eff}^{(s)}&=\int d^4x\bar{\Psi}(x)[i\Gamma^\mu\partial_\mu-\alpha\Gamma^1\Gamma^5\tau^z+(k_0-s\partial_3\theta_s)\Gamma^3\Gamma^5s^z]\Psi(x)\nonumber \\
&+i\int d^4x\int_0^sds \frac{\theta_s}{4\pi^2}\epsilon^{\mu\nu\alpha\beta}F_{s,\mu\nu}F_{\alpha\beta} .
\end{align}
Setting $\theta_s=k_0x^3$ and $s=1$, we obtain
\begin{gather}
S_{eff}^{(1)}=\int d^4x\bar{\Psi}(x)[i\Gamma^\mu\partial_\mu-\alpha\Gamma^1\Gamma^5\tau^z]\Psi(x)+i\int d^4x\int_0^sds \frac{k_0x^3}{4\pi^2}\epsilon^{\mu\nu\alpha\beta}F_{s,\mu\nu}F_{\alpha\beta},
\end{gather}
where the topological-in-origin $\theta-$term is given by
\begin{gather}
S_{eff}^{s}\equiv \int d^4x \frac{ik_0 x^3}{4\pi^2}\epsilon^{\mu\nu\alpha\beta}F_{s,\mu\nu}F_{\alpha\beta} .
\end{gather}

\section{Derivation of a pair of Fermi-arc surface states for inversion symmetry-breaking Weyl metals} \label{Appendix:Boundary-Mode-IB-Weyl}

\subsection{Valley Hall current}

If we express the Hamiltonian in terms of $\gamma_v^\mu$ matrices, we have
\begin{align}
H_{WM}&=\int d^3x\bar{\Psi}(x)\Big[i\gamma_v^1\partial_1+i\gamma_v^2\partial_2+i\gamma_v^3\tau^z\partial_3-\alpha\gamma_v^1\gamma_v^5-k_0\gamma_v^3+m\tau^z\Big]\Psi(x)\nonumber \\
&=\int d^3x\Psi^\dagger(x)\Big[i\gamma_v^0\gamma_v^1\partial_1+i\gamma_v^0\gamma_v^2\partial_2+i\gamma_v^0\gamma_v^3\tau^z\partial_3-\alpha\gamma_v^0\gamma_v^1\gamma_v^5-k_0\gamma_v^0\gamma_v^3+m\gamma_v^0\tau^z\Big]\Psi(x) ,
\end{align}
where
\begin{gather}
\gamma_v^0=s^x\sigma^x, \; \gamma_v^1=-is^y,\; \gamma_v^2=-is^x\sigma^z,\; \gamma_v^3=-is^x\sigma^y,\; \gamma_v^5=-s^z.
\end{gather}
Here, we introduced a term $m\tau^z\gamma_v^0\Psi^\dagger\Psi$ that gives a mass to each valley. Since this mass term preserves the time reversal symmetry, it is allowed.

The above Hamiltonian looks quite similar to the Hamiltonian of the time-reversal symmetry-broken Weyl metal except for the representation of gamma matrices. In this respect we may use the boundary solution of the time-reversal symmetry-broken Weyl metal in order to find that of the inversion symmetry-broken Weyl metal. Unfortunately, it is not much straightforward to apply the case of the time-reversal symmetry-breaking to that of the inversion symmetry-breaking directly since the representations of gamma matrices are different from each other. Therefore, we perform the canonical transformation to change the representation into Weyl one. Since the canonical transformation in the particle-number conserving system is nothing but the unitary transformation, we have
\begin{gather}
\mathcal{H}=\Psi^\dagger H\Psi=\Psi^\dagger U^\dagger UHU^\dagger U\Psi\equiv \Psi'^\dagger H'\Psi' \\
\Psi'=U\Psi,\; H'=UHU^\dagger .
\end{gather}
Here, unitary matrix $U$ should satisfy the following relations
\begin{gather}
U\gamma_v^0\gamma_v^1U^\dagger=\gamma^0\gamma^1,\; U\gamma_v^0\gamma_v^2U^\dagger=\gamma^0\gamma^2\\
U\gamma_v^0\gamma_v^3U^\dagger=\gamma^0\gamma^3,\; U\gamma_v^0\gamma_v^5U^\dagger=\gamma^0\gamma^5 ,
\end{gather}
where $\gamma^0=\Big(\begin{array}{cc} 0 & 1 \\ 1 & 0 \end{array}\Big)$, $\gamma^k=\Big(\begin{array}{cc} 0 & \sigma^k \\ -\sigma^k & 0 \end{array} \Big)$ and $\gamma^5=i\gamma^0\gamma^1\gamma^2\gamma^3$. Then, the resulting unitary matrix $U$ is given by
\begin{gather}
U=\eta \sigma^2\otimes \frac{1+s^3}{2}+\xi \sigma^3\otimes \frac{1-s^3}{2}=\left(\begin{array}{cc}\eta \sigma^2 & 0 \\ 0 & \xi \sigma^3
\end{array}\right),\; \Big\{\begin{array}{c}\eta^*\eta=1 \\ \xi^*\xi=1\end{array} ,
\end{gather}
where $\eta\xi^*=i$ ans $U\gamma_v^i U^\dagger=\gamma^i$ are satisfied.

Under this canonical transformation, we note that both time reversal symmetry and inversion symmetry operators are changed as well
\begin{gather}
\tilde{P}=\sigma^z\otimes\tau^x\rightarrow \bar{P}=U\sigma^z\otimes \tau^x U^\dagger = -\sigma^z\otimes s^z\otimes \tau^z\\
\tilde{T}=is^y\otimes \tau^x \mathcal{K}\rightarrow \bar{T}=U is^y\otimes \tau^x \mathcal{K}U^\dagger=-\eta\xi \sigma^x\otimes s^y\otimes \tau^x\mathcal{K} .
\end{gather}

Rewriting the Hamiltonian in terms of this Weyl representation of gamma matrices, we obtain
\begin{align}
H_{WM}=\int d^3x\Psi'^\dagger(x)\Big[i\gamma^0\gamma^1\partial_1+i\gamma^0\gamma^2\partial_2+i\gamma^0\gamma^3\tau^z\partial_3-\alpha\gamma^0\gamma^1\gamma^5-k_0\gamma^0\gamma^3+m\gamma^0\tau^z\Big]\Psi'(x) ,
\end{align}
where $\Psi'(x)=U\Psi(x)$. Since we do not take into account any scattering terms between different valleys ($\tau^z=1$ and $\tau^z=-1$), $\tau^z$ must be a good quantum number. As a result, we can divide the above Hamiltonian into two sectors of $\tau^z=1$ and $\tau^z=-1$, given by
 \begin{gather}
H_{WM}=H_{WM,\tau^z=1}+H_{WM,\tau^z=-1}\\
H_{WM,\tau^z=\pm1}=\int d^3x\Psi_{\pm}'^\dagger\Big[i\gamma^0\gamma^1\partial_1+i\gamma^0\gamma^2\partial_2\pm i\gamma^0\gamma^3\partial_3-\alpha\theta(z)\gamma^0\gamma^1\gamma^5-k_0\gamma^0\gamma^3\pm m\gamma^0\Big]\Psi'_{\pm}\\
\Psi'_{\pm}=\Psi'_{\tau^z=\pm 1},\; \Psi'=\left(\begin{array}{c}\Psi'_{\tau^z=+1} \\ \Psi'_{\tau^z=-1}\end{array}\right) .
\end{gather}
In the above we set $\alpha=\alpha\theta(z)$ to get a boundary solution at the $z=0$ plane. Then, the Dirac equation is also separated and given by
\begin{gather}
H_{surf}\psi(x,y,z)=E\psi(x,y,z),\; H_{surf}=H_{surf,\tau^z=1}\oplus H_{surf,\tau^z=-1}\\
\Rightarrow \psi(x,y,z)=\Big(\begin{array}{c}\psi_{\tau^z=1} \\ \psi_{\tau^z=-1} \end{array}\Big),\; \Big\{\begin{array}{c} H_{surf,\tau^z=1}\psi_{\tau^z=1}=E_{+}\psi_{\tau^z=1},\\ H_{surf,\tau^z=-1}\psi_{\tau^z=-1}=E_{-}\psi_{\tau^z=-1}\end{array} .
\end{gather}

First, we consider
\begin{enumerate}
\item[(i)] $\tau^z=1$
\begin{gather}
H_{surf,\tau^z=1}\psi_{\tau^z=1}=E_{+}\psi_{\tau^z=1}\\
\left(\begin{array}{cc}-i\vec{\sigma}\cdot\nabla-\alpha\theta(z)\sigma^1+k_0\sigma^3 & m \\ m & i\vec{\sigma}\cdot\nabla-\alpha\theta(z)\sigma^1-k_0\sigma^3 \end{array}\right)
\psi_{\tau^z=1}(x,y,z)=E_{+}\psi_{\tau^z=1}(x,y,z) .
\end{gather}
Since there are translational symmetries along the $x-$ and $y-$ axis, we set $\psi_{\tau^z=1}(x,y,z)=e^{ik_x x+ik_y y}\phi_{\tau^z=1,k_x,k_y}(z)$ and obtain
\begin{align}
&\left(\begin{array}{cc}\sigma^1(k_x-\alpha\theta(z))+\sigma^2k_y+\sigma^3(-i\partial_z+k_0) & m \\ m & -\sigma^1(k_x+\alpha\theta(z))-\sigma^2k_y+\sigma^3(i\partial_z-k_0)\end{array}\right)\nonumber \\
&\times\phi_{\tau^z=1,k_x,k_y}(z)=E_+\phi_{\tau^z=1,k_x,k_y}(z) .
\end{align}

In order to solve this equation, we use the following ansatz
\begin{gather}
\phi_{+,k_x,k_y}(z)=u_{+}(z)\left(\begin{array}{c}1 \\ i \\ 0 \\ 0 \end{array}\right)+v_{+}(z)\left(\begin{array}{c} 0\\ 0 \\ 1\\ -i\end{array}\right) .
\end{gather}
We note that both eigenstates have the eigenvalue $-1$ for $\gamma^0\gamma^2$. Then, we obtain
\begin{gather}
\Big(ik_x-(i\alpha\theta(z)+i\partial_z-k_0)\Big)u_+(z)+mv_+(z)=0\\
mu_+(z)+\Big(ik_x+(i\alpha\theta(z)+i\partial_z-k_0)\Big)v_+(z)=0\\
E_+=k_y\\
\Rightarrow \Big\{\begin{array}{c}i(k_x-\partial_z)\tilde{u}_+(z)+m\tilde{v}_+(z)=0\\ m\tilde{u}_+(z)+i(k_x+\partial_z)\tilde{v}_+(z)=0\end{array},\; \Big\{\begin{array}{c}u_{+}(z)=e^{-ik_0z-\alpha\theta(z)z}\tilde{u}_+(z) \\v_{+}(z)=e^{-ik_0z-\alpha\theta(z)z}\tilde{v}_+(z)
\end{array}\\
\Rightarrow (\partial_z^2-(k_x^2+m^2))\tilde{u}_+(z)/\tilde{v}_{+}(z)=0\\
\Rightarrow \Big\{\begin{array}{c}\tilde{u}_+(z)=A_+^1e^{-\sqrt{k_x^2+m^2}z}+A_+^2e^{\sqrt{k_x^2+m^2}z}\\ \tilde{v}_+(z)=B_+^1e^{-\sqrt{k_x^2+m^2}z}+B_+^2e^{\sqrt{k_x^2+m^2}z}\end{array} .
\end{gather}

Considering the boundary conditions and normalizability of the wave functions as discussed in the case of time-reversal symmetry-breaking, we find
\begin{gather}
-\sqrt{\alpha^2-m^2}<k_x<\sqrt{\alpha^2-m^2}\\
\phi_{+,k_x,k_y}(z)=A_+^2e^{-ik_0z}e^{-\alpha\theta(z)z}e^{\sqrt{k_x^2+m^2}z}\left(\begin{array}{c}1\\ i \\ i\frac{m}{\sqrt{k_x^2+m^2}+k_x} \\ \frac{m}{\sqrt{k_x^2+m^2}+k_x}\end{array}\right)\\
 |A_+^2|^2=\frac{m^2}{m^2+(\sqrt{k_x^2+m^2}-k_x)^2}\frac{\sqrt{k_x^2+m^2}(\alpha-\sqrt{k_x^2+m^2})}{\alpha},\\
E_+=k_y .
\end{gather}

Next, we consider
\item[(ii)] $\tau^z=-1$
\begin{gather}
H_{surf,\tau^z=-1}\psi_{\tau^z=-1}=E_-\psi_{\tau^z=-1}\\
\left(\begin{array}{cc}-i\vec{\sigma}\cdot\nabla_\perp+i\sigma^3\partial_z-\alpha\theta(z)\sigma^1+k_0\sigma^3 & -m \\ -m & i\vec{\sigma}\cdot\nabla_\perp-i\sigma^3\partial_z-\alpha\theta(z)\sigma^1-k_0\sigma^3 \end{array}\right)\psi_{\tau^z=-1}(x,y,z)\nonumber \\
=E_-\psi_{\tau^z=-1}(x,y,z) .
\end{gather}
Setting $\psi_{\tau^z=-1}(x,y,z)=e^{ik_x x+ik_y y}\phi_{\tau^z=-1,k_x,k_y}(z)$, we have
\begin{align}
&\left(\begin{array}{cc}\sigma^1(k_x-\alpha\theta(z))+\sigma^2k_y+\sigma^3(i\partial_z+k_0) & -m \\ -m & -\sigma^1(k_x+\alpha\theta(z))-\sigma^2k_y+\sigma^3(-i\partial_z-k_0)\end{array}\right)\nonumber \\
&\times\phi_{\tau^z=-1,k_x,k_y}(z)=E_-\phi_{\tau^z=-1,k_x,k_y}(z) .
\end{align}

Now, we consider the ansatz of
\begin{gather}
\phi_{\tau^z=-1,k_x,k_y}(z)=u_-(z)\left(\begin{array}{c} 1\\ -i\\0
\\0\end{array}\right)+v_-(z)\left(\begin{array}{c}0\\ 0 \\ 1\\ i \end{array}\right) ,
\end{gather}
where both eigenstates have the eigenvalue $1$ for $\gamma^0\gamma^2$. Then, we obtain
\begin{gather}
\Big(i\partial_z+k_0+i\alpha\theta(z)-ik_x\Big)u_-(z)-mv_-(z)=0\\
mu_-(z)+\Big(i\partial_z+k_0+i\alpha\theta(z)+ik_x\Big)v_-(z)=0\\
E_-=-k_y\\
\Rightarrow \Big\{\begin{array}{c}i(\partial_z-k_x)\tilde{u}_-(z)-m\tilde{v}_-(z)=0 \\ m\tilde{u}_-+i(\partial_z+k_x)\tilde{v}_-(z)=0\end{array},\; \Big\{\begin{array}{c}u_(z)= e^{ik_0 z-\alpha\theta(z)z}\tilde{u}_-(z)\\v_(z)= e^{ik_0 z-\alpha\theta(z)z}\tilde{v}_-(z) \end{array}\\
\Rightarrow (\partial_z^2-(k_x^2+m^2))\tilde{u}_-(z)/\tilde{v}_-(z)=0\\
\Rightarrow \Big\{\begin{array}{c}\tilde{u}_-(z)=A_-^1e^{-\sqrt{k_x^2+m^2}z}+A_-^2e^{\sqrt{k_x^2+m^2}z}\\ \tilde{v}_-(z)=B_-^1e^{-\sqrt{k_x^2+m^2}z}+B_-^2e^{\sqrt{k_x^2+m^2}z}\end{array} .
\end{gather}

Considering the boundary conditions and normalizability of the wave functions, we obtain
\begin{gather}
-\sqrt{\alpha^2-m^2}<k_x<\sqrt{\alpha^2-m^2}\\
\phi_{-,k_x,k_y}(z)=A_-^2e^{ik_0z}e^{-\alpha\theta(z)z}e^{\sqrt{k_x^2+m^2}z}\left(\begin{array}{c}1\\ -i \\ i\frac{m}{\sqrt{k_x^2+m^2}+k_x} \\ -\frac{m}{\sqrt{k_x^2+m^2}+k_x}\end{array}\right)\\
 |A_-^2|^2=\frac{m^2}{m^2+(\sqrt{k_x^2+m^2}-k_x)^2}\frac{\sqrt{k_x^2+m^2}(\alpha-\sqrt{k_x^2+m^2})}{\alpha},\\
E_-=-k_y .
\end{gather}

In summary, we find a pair of Fermi-arc surface states
\begin{gather}
-\sqrt{\alpha^2-m^2}<k_x<\sqrt{\alpha^2-m^2}\\
\psi_{\tau^z=1,k_x,k_y}=A(k_x)e^{ik_x x}e^{ik_y y}e^{-ik_0 z}e^{-\alpha\theta(z)z}e^{\sqrt{k_x^2+m^2}z}\left(\begin{array}{c}1\\ i\\ C(k_x) \\ -iC(k_x)\end{array}\right),\; E_{+}=k_y\\
\psi_{\tau^z=-1,k_x,k_y}=A(k_x)e^{ik_x x}e^{ik_y y}e^{ik_0 z}e^{-\alpha\theta(z)z}e^{\sqrt{k_x^2+m^2}z}\left(
\begin{array}{c} 1 \\ -i \\ C(k_x) \\ iC(k_x)\end{array}\right),\; E_-=-k_y\\
|A(k_x)|^2=\frac{m^2}{m^2+(\sqrt{k_x^2+m^2}-k_x)^2}\frac{\sqrt{k_x^2+m^2}(\alpha-\sqrt{k_x^2+m^2})}{\alpha}\\
C(k_x)=i\frac{m}{\sqrt{k_x^2+m^2}+k_x} ,
\end{gather}
where they are characterized by opposite chirality quantum numbers given by
\begin{gather}
\gamma^0\gamma^2\psi_{\tau^z=1,k_x,k_y}=-\psi_{\tau^z=1,k_x,k_y},\;\gamma^0\gamma^2\psi_{\tau^z=-1,k_x,k_y}=+\psi_{\tau^z=-1,k_x,k_y} .
\end{gather}
\end{enumerate}

\subsection{Spin Hall current}

The effective Hamiltonian in the $\gamma_s^\mu$ representation is
\begin{align}
H_{WM}=\int d^3 x \Psi^\dagger(x)\Big[i\gamma_s^0\gamma_s^1 s^z\partial_x+i\gamma_s^0\gamma_s^2 \partial_y+i\gamma_s^0\gamma_s^3\partial_z+\alpha \gamma_s^0 \gamma_s^1+k_0\gamma_s^0\gamma_s^3\gamma_s^5+m\gamma_s^0\Big]\Psi(x) ,
\end{align}
where
\begin{gather}
\gamma_s^0=\sigma^z\tau^x,\; \gamma_s^1=i\sigma^y\tau^x,\; \gamma_s^2=i\sigma^x\tau^x,\; \gamma_s^3=-i\tau^y,\; \gamma_s^5=-\tau^z.
\end{gather}
Here, we also introduced a term $m\gamma_s^0\Psi^\dagger\Psi$ that gives a mass to to each spin sector. This mass term also respects the time reversal symmetry.

Following the previous section, we find the canonical transformation
\begin{gather}
U=\eta \sigma^2\otimes \frac{1+\tau^3}{2}+\xi \sigma^1\otimes \frac{1-\tau^3}{2}=\left(\begin{array}{cc} \eta \sigma^2 & 0 \\ 0 & \xi \sigma^1\end{array}\right),\; \eta\eta^*=\xi\xi^*=1 ,
\end{gather}
where $\eta^*\xi=i$ and $U\gamma_s^i U^\dagger=\gamma^i$ are satisfied.

Under this canonical transformation, both the time reversal symmetry and inversion symmetry operators are also changed as follows
\begin{gather}
\tilde{P}=\sigma^z\otimes \tau^x\rightarrow \bar{P}=U\sigma^z\otimes \tau^xU^\dagger=\tau^x\\
\tilde{T}=i s^y\otimes \tau^x\mathcal{K}\rightarrow \bar{T}=U is^y\otimes \tau^x \mathcal{K}U^\dagger=\eta\xi s^y\otimes \sigma^z\otimes \tau^x \mathcal{K} .
\end{gather}

Now, we start from
\begin{gather}
H_{WM}=\int d^3 x \Psi'^\dagger(x)\Big[i\gamma^0\gamma^1 s^z \partial_x + i\gamma^0\gamma^2\partial_y+i\gamma^0\gamma^3\partial_z+\alpha\gamma^0\gamma^1+k_0\gamma^0\gamma^3\gamma^5+m\gamma^0\Big]\Psi'(x) ,
\end{gather}
where $\Psi'(x)=U\Psi(x)$. Since we do not consider any scattering terms between different spin sections ($s^z=1$ and $s^z=-1$), $s^z$ is a good quantum number. As a result, the above Hamiltonian is separated into two spin sectors with $s^z=1$ and $s^z=-1$
\begin{gather}
H_{WM}=H_{WM,s^z=1}+H_{WM,s^z=-1}\\
H_{WM,s^z=\pm1}=\int d^3 x\Psi'^\dagger_{\pm}\Big[\pm i\gamma^0\gamma^1  \partial_x + i\gamma^0\gamma^2\partial_y+i\gamma^0\gamma^3\partial_z+\alpha\gamma^0\gamma^1+k_0\theta(x)\gamma^0\gamma^3\gamma^5+m\gamma^0\Big]\Psi_\pm'\\
\Psi'_{\pm}=\Psi'_{s^z=\pm 1},\; \Psi'=\left(\begin{array}{c}\Psi'_{s^z=+1} \\ \Psi'_{s^z=-1} \end{array}\right).
\end{gather}
Here, we set $k_0=k_0\theta(x)$ to get a boundary solution at the $x=0$ plane. Accordingly, the Dirac equation is
\begin{gather}
H_{surf}\psi(x,y,z)=E\psi(x,y,z),\; H_{surf}=H_{surf,s^z=1}\oplus H_{surf,s^z=-1}\\
\Rightarrow \psi(x,y,z)=\left(\begin{array}{c}\psi_{s^z=1}\\ \psi_{s^z=-1}\end{array}\right),\; \Big\{\begin{array}{c}H_{surf,s^z=1}\psi_{s^z=1}=E_+\psi_{s^z=1}, \\ H_{surf,s^z=-1}\psi_{s^z=-1}=E_-\psi_{s^z=-1}\end{array} .
\end{gather}

First, we consider
\begin{itemize}
\item[(i)] $s^z=1$
\begin{gather}
H_{surf,s^z=1}\psi_{s^z=1}=E_+\psi_{s^z=1}\\
\left(\begin{array}{cc}-i\vec{\sigma}\cdot\nabla-\alpha\sigma^1+k_0\theta(x)\sigma^3 & m \\ m & i\vec{\sigma}\cdot\nabla+\alpha\sigma^1+k_0\theta(x)\sigma^3\end{array}\right)\psi_{s^z=1}(x,y,z)=E_+\psi_{s^z=1}(x,y,z) .
\end{gather}
Since there are translational symmetries along the $y-$ and $z-$ axis, we set $\psi_{s^z=1}(x,y,z)=e^{ik_y y+ik_z z}\phi_{s^z=1,k_y,k_z}(x)$ and obtain
\begin{align}
&\left(\begin{array}{cc}\sigma^1(-i\partial_x-\alpha)+\sigma^2k_y+\sigma^3(k_z+k_0\theta(x)) & m \\ m & \sigma^1(i\partial_x+\alpha)-\sigma^2 k_y-\sigma^3(k_z-k_0\theta(x))\end{array}\right)\nonumber \\
&\times \phi_{s^z=1,k_y,k_z}(x)=E_+ \phi_{s^z=1,k_y,k_z}(x) .
\end{align}

Following the previous section, we use the ansatz of
\begin{gather}
\phi_{+,k_y,k_z}(x)=u_+(x)\left(\begin{array}{c}1 \\ i \\ 0\\ 0\end{array}\right)+v_+(x)\left(\begin{array}{c}0 \\ 0 \\ 1 \\ -i\end{array}\right) .
\end{gather}
Here, both eigenstates have the eigenvalue $\gamma^0\gamma^2=-1$. Then, we obtain
\begin{gather}
(\partial_x-i\alpha+k_0\theta(x)+k_z)u_+(x)+mv_+(x)=0\\
mu_+(x)+(\partial_x-i\alpha+k_0\theta(x)-k_z)v_+(x)=0\\
E_+=k_y\\
\Rightarrow \Big\{\begin{array}{c}(\partial_x+k_z)\tilde{u}_+(x)+m\tilde{v}_+(x)=0 \\ m\tilde{u}_+(x)+(\partial_x-k_z)\tilde{v}_+(x)=0\end{array}, \; \Big\{\begin{array}{c}u_+(x)=e^{i\alpha x-k_0\theta(x)x}\tilde{u}_+(x) \\ v_+(x)=e^{i\alpha x-k_0\theta(x)x}\tilde{v}_+(x)\end{array}\\
\Rightarrow (\partial_x^2-(k_z^2+m^2))\tilde{u}_+(x)/\tilde{v}_+(x)=0\\
\Rightarrow \Big\{\begin{array}{c}\tilde{u}_+(x)=A_+^1e^{-\sqrt{k_z^2+m^2}x}+A_+^2 e^{\sqrt{k_z^2+m^2}x} \\ \tilde{v}_+(x)=B_+^1e^{-\sqrt{k_z^2+m^2}x}+B_+^2 e^{\sqrt{k_z^2+m^2}x}\end{array} .
\end{gather}
Considering the boundary conditions and normalizability of the wave functions as discussed in the previous section, we find
\begin{gather}
-\sqrt{k_0^2-m^2}<k_z<\sqrt{k_0^2-m^2}\\
\phi_{+,k_y,k_z}=A_+^2 e^{i\alpha x}e^{-k_0\theta(x)x}e^{\sqrt{k_z^2+m^2}}\left(\begin{array}{c}1 \\ i \\ -\frac{m}{\sqrt{k_z^2+m^2}-k_z}\\ \frac{im}{\sqrt{k_z^2+m^2}-k_z}\end{array}\right)\\
E_+=k_y .
\end{gather}
Here, $A_+^2$ is the same as that of the previous section.

Next, we consider
\item[(ii)] $s^z=-1$
\begin{gather}
H_{surf,s^z=-1}\psi_{s^z=-1}=E_- \psi_{s^z=-1}\\
\left(\begin{array}{cc}i\sigma^1\partial_x -i\vec{\sigma}\cdot\nabla_\perp-\alpha\sigma^1+k_0\theta(x)\sigma^3 & m \\ m & -i\sigma^1\partial_x+i\vec{\sigma}\cdot\nabla_\perp +\alpha\sigma^1+k_0\theta(x)\sigma^3 \end{array}\right)\psi_{s^z=-1}(x,y,z)\nonumber \\
=E_-\psi_{s^z=-1}(x,y,z) .
\end{gather}
Taking $\psi_{s^z=-1}(x,y,z)=e^{ik_y y+ik_z z}\phi_{s^z=-1,k_y,k_z}$, we have
\begin{align}
&\left(\begin{array}{cc}\sigma^1(i\partial_x-\alpha)+\sigma^2k_y+\sigma^3(k_z+k_0\theta(x)) & m \\ m & \sigma^1(-i\partial_x+\alpha)-\sigma^2 k_y-\sigma^3(k_z-k_0\theta(x))\end{array}\right)\nonumber \\
&\times \phi_{s^z=-1,k_y,k_z}(x)=E_- \phi_{s^z=-1,k_y,k_z}(x) .
\end{align}
Now, the ansatz is
\begin{gather}
\phi_{s^z=-1,k_y,k_z}(x)=u_-(x)\left(\begin{array}{c}1 \\ -i \\ 0 \\ 0 \end{array}\right)+v_-(x)\left(\begin{array}{c}0 \\ 0 \\ 1\\ i \end{array}\right) ,
\end{gather}
where both eigenstates have the eigenvalue $\gamma^0\gamma^2=1$. As a result, we obtain
\begin{gather}
(\partial_x+i\alpha+k_0\theta(x)+k_z)u_-(x)+mv_-(x)=0\\
mu_-(x)+(\partial_x+i\alpha+k_0\theta(x)-k_z)=0\\
E_-=-k_y\\
\Rightarrow \Big\{\begin{array}{c}(\partial_x+k_z)\tilde{u}_-(x)+m\tilde{v}_-(x)=0 \\ m\tilde{u}_-(x)+(\partial_x-k_z)\tilde{v}_-(x)=0 \end{array},\; \Big\{\begin{array}{c}u_-(x)=e^{-i\alpha x-k_0\theta(x) x}\tilde{u}_-(x) \\ v_-(x)=e^{-i\alpha x-k_0\theta(x)x}\tilde{v}_-(x)\end{array}\\
\Rightarrow (\partial_x^2-(k_z^2+m^2))\tilde{u}_-(x)/\tilde{v}_-(x)=0\\
\Rightarrow \Big\{\begin{array}{c}\tilde{u}_-(x)=A_-^1e^{-\sqrt{k_z^2+m^2}x}+A_-^2e^{\sqrt{k_z^2+m^2}x}\\ \tilde{v}_-(x)=B_-^1e^{-\sqrt{k_z^2+m^2}x}+B_-^2e^{\sqrt{k_z^2+m^2}x}\end{array} .
\end{gather}
Considering the boundary conditions and normalizability of the wave functions as discussed in the previous section, we find
\begin{gather}
-\sqrt{k_0^2-m^2}<k_z<\sqrt{k_0^2-m^2}\\
\phi_{-,k_y,k_z}=A_-^2e^{i\alpha x}e^{-k_0 \theta(x)x}e^{\sqrt{k_z^2+m^2}x}\left(\begin{array}{c}1 \\ -i \\ -\frac{m}{\sqrt{k_z^2+m^2}-k_z}\\ -\frac{im}{\sqrt{k_z^2+m^2}-k_z}\end{array}\right) .
\end{gather}
Here, $A_-^2$ is the same as that of the previous section.

In summary, we find a pair of Fermi-arc surface states
\begin{gather}
-\sqrt{k_0^2-m^2}<k_z<\sqrt{k_0^2-m^2}\\
\psi_{s^z=1,k_y,k_z}=A(k_z)e^{ik_y y+ik_z z}e^{i\alpha x}e^{-k_0\theta(x)x}e^{\sqrt{k_z^2+m^2}x}\left(\begin{array}{c}1 \\ i \\ C(k_z) \\ -iC(k_z)\end{array}\right), \; E_+=k_y\\
\psi_{s^z=-1,k_y,k_z}=A(k_z)e^{ik_y y+ik_z z}e^{i\alpha x}e^{-k_0\theta(x)x}e^{\sqrt{k_z^2+m^2}x}\left(\begin{array}{c}1 \\ -i \\ -C(k_z) \\ -i C(k_z)\end{array}\right),\; E_-=-k_y\\
C(k_z)=\frac{m}{\sqrt{k_z^2+m^2}-k_z} ,
\end{gather}
where they are characterized by opposite chirality quantum numbers given by
\begin{gather}
\gamma^0\gamma^2\psi_{s^z=1,k_y,k_z}=-\psi_{s^z=1,k_y,k_z},\; \gamma^0\gamma^2\psi_{s^z=-1,k_y,k_z}=+\psi_{s^z=-1,k_y,k_z} .
\end{gather}
\end{itemize}

\section{Calculation of the one-loop quantum correction to the U(1) surface current in the case of inversion symmetry-breaking Weyl metals} \label{Appendix:1-loop-IB-Weyl}

\subsection{Valley Hall current}

Since we do not take into account any interactions between fields with different $k_x$ momentum, we will not include the summation over $k_x$ from now on. Introducing the Pauli-Villars regularization field into the effective action, we start from the following surface action
\begin{gather}
S=\int d^2x \Big[\bar{\Psi}(x)i\gamma^\mu(\partial_\mu+iA_\mu+iV_\mu\bar{\gamma})\Psi(x)+\bar{\phi}(x)i\gamma^\mu(\partial_\mu+iA_\mu+iV_\mu\bar{\gamma})\phi(x)+\bar{\phi}(x)M\phi(x)\Big] .
\end{gather}
If we consider the valley gauge transformation of $\bar{\phi}(x)$ and $\phi(x)$ like that of $\bar{\Psi}(x)$ and $\Psi(x)$, the mass term breaks the valley gauge symmetry explicitly.

Under the Fourier transformations
\begin{gather}
\Psi(x)=\int \frac{d^2k}{(2\pi)^2}e^{-ik\cdot x}\Psi(x),\;\; \bar{\Psi}(x)=\int \frac{d^2k}{(2\pi)^2}e^{ik \cdot x}\bar{\Psi}(k),\\
\phi(x)=\int \frac{d^2k}{(2\pi)^2}e^{-ik\cdot x}\phi(x),\;\; \bar{\phi}(x)=\int \frac{d^2k}{(2\pi)^2}e^{ik \cdot x}\bar{\phi}(k)\\
A_\mu(q)=\int d^2x e^{iq\cdot x}A_\mu(x),\; V_\mu(q)=\int d^2x e^{iq\cdot x}A_\mu(x) ,
\end{gather}
we obtain
\begin{align}
S&=\int \frac{d^2k}{(2\pi)^2}\Big[\bar{\Psi}(k)\slashed{k}\Psi(k)+\bar{\phi}(k)(\slashed{k}+M)\phi(k) \nonumber \\&-\int \frac{d^2q}{(2\pi)^2}\Big(\bar{\Psi}(k+q)\slashed{A}(q)\mathcal{P}_+\Psi(k)+\bar{\phi}(k+q)\slashed{A}(q)\mathcal{P}_+\phi(k)\Big)\nonumber \\
&-\int\frac{d^2q}{(2\pi)^2}\Big(\bar{\Psi}(k+q)\slashed{V}(q)\bar{\gamma}\Psi(k)+\bar{\phi}(k+q)\slashed{V}(q)\bar{\gamma}\phi(k)\Big)\Big] ,
\end{align}
where both Green's functions are
\begin{gather}
G(k)=\langle \Psi(k)\bar{\Psi}(k)\rangle=\frac{\slashed{k}}{k^2}, \;\;  \tilde{G}(k)= \langle \phi(k)\bar{\phi}(k)\rangle=\frac{\slashed{k}-M}{k^2-M^2} .
\end{gather}

The valley current of
\begin{align}
j_v^\mu(q)&=\int d^2x e^{-iq\cdot x}j_v^\mu(x)=\int d^2xe^{-iq\cdot x}\bar{\Psi}(x)\gamma^\mu\bar{\gamma}\Psi(x) =\int \frac{d^2k}{(2\pi)^2}\bar{\Psi}(k+q)\gamma^\mu\bar{\gamma}\Psi(k)
\end{align}
is regularized as
\begin{gather}
j_{reg}^\mu(q)=\int \frac{d^2k}{(2\pi)^2}\Big[\bar{\Psi}(k+q)\gamma^\mu \bar{\gamma} \Psi(k)+\bar{\phi}(k+q)\gamma^\mu \bar{\gamma}\phi(k)\Big] .
\end{gather}

Up to the one-loop order, there are two contributions from $A_{\mu}$ and $V_\mu$, respectively, in the following way
\begin{align}
\langle j_{reg}^{\mu}(q)\rangle&=\int \frac{d^2k}{(2\pi)^2}\Big[-tr\Big(G(k)\gamma^\mu \bar{\gamma}G(k+q)\gamma^\nu\Big)+tr\Big(\tilde{G}(k)\gamma^\mu\bar{\gamma}\tilde{G}(k+q)\gamma^\nu\Big)\Big]A_{\nu}(-q)\nonumber \\
&+\int \frac{d^2k}{(2\pi)^2}\Big[tr\Big(G(k)\gamma^\mu\bar{\gamma}G(k+q)\gamma^\nu\bar{\gamma}\Big)-tr\Big(\tilde{G}(k)\gamma^\mu\bar{\gamma}\tilde{G}(k+q)\gamma^\nu\bar{\gamma}\Big)\Big]V_{\nu}(-q).
\end{align}
As a result, we find
\begin{align}
\therefore \langle j_{reg}^{\mu}(q)\rangle&=\sum_{\nu}\Big[\frac{i\Big(\epsilon^{\mu\nu}(q_\mu^2-q_\nu^2)+2\delta_{\mu\nu}\sum_{\alpha}\epsilon^{\alpha\mu}q_\mu q_\alpha\Big)}{2\pi q^2}-\frac{i}{2\pi}\epsilon^{\mu\nu}\Big]A_{\nu}(-q)\nonumber \\
&+\sum_{\nu}\frac{2q^\mu q^\nu}{\pi q^2}V_{\nu}(-q)\\
\Rightarrow  q_\mu \langle j_{reg}^\mu(q)\rangle &=\sum_{\mu,\nu}\Big[\frac{i}{2\pi q^2}\Big(\epsilon^{\mu\nu}q_\mu(q_\mu^2-q_\nu^2)+2q_\mu \delta_{\mu\nu}\sum_{\alpha}\epsilon^{\alpha\mu}q_\mu q_\alpha\Big)
-\frac{i}{2\pi}\epsilon^{\mu\nu}q_\mu\Big]A_\nu(-q)\nonumber \\
&+\sum_{\nu}\frac{2}{\pi}q^\nu V_{\nu}(-q)\nonumber \\
&=\frac{i}{2\pi}\sum_{\mu,\nu}\Big[\epsilon^{\mu\nu}q_\mu(q_\mu^2+q_\nu^2)/q^2-\epsilon^{\mu\nu}q_\mu\Big]A_{\nu}(-q)+\sum_{\nu}\frac{2}{\pi}q^\nu V_\nu(-q)\nonumber \\
&=-\frac{i}{\pi}\sum_{\mu,\nu}\epsilon^{\mu\nu}q_\mu A_\nu(-q)+\sum_{\nu}\frac{2}{\pi}q^\nu V_\nu(-q) .
\end{align}

If we ignore the contribution from the $V_\nu$ field considering that it is a fictitious gauge field, we obtain the anomaly
\begin{gather}
q_\mu \langle j_{reg}^\mu(q)\rangle=-\frac{i}{\pi}\sum_{\mu,\nu}\epsilon^{\mu\nu}q_\mu A_\nu(-q)\\
\Rightarrow \partial_\mu \langle j_{reg}^\mu(x)\rangle =-\frac{i}{\pi}\epsilon^{\mu\nu}\partial_\mu A_\nu (x)=-\frac{i}{2\pi}\epsilon^{\mu\nu}F_{\mu\nu}(x) .
\end{gather}
One can show that this anomaly is canceled by the anomaly inflow of the bulk, considering
\begin{align}
\delta_\eta W[V,A]
& \equiv W[\mathcal{V}+d\eta,\mathcal{A}]-W[\mathcal{V},\mathcal{A}]=\int d^2x \partial_\mu \eta(x)\frac{\delta W}{\delta V_\mu}\nonumber \\
&=\int d^2x\partial_\mu\eta(x)j_{v}^{\mu}(x)=-\int d^2x \eta(x)\partial_\mu j_v^{\mu}(x)\nonumber \\
&=\frac{i}{2\pi}\int d^2x \eta(x)\epsilon^{\mu\nu}F_{\mu\nu}(x) .
\end{align}

\subsection{Spin Hall current}

The surface Hamiltonian in terms of the $|s^z,k_y,k_z\rangle$ basis is given by
\begin{align}
H&=\sum_{-\sqrt{k_0^2-m^2}<k_z<\sqrt{\alpha^2-m^2}}\sum_{k_y}\Psi^\dagger(k_y,k_z)\left(\begin{array}{cc}k_y & 0 \\ 0 & -k_y \end{array}\right)\Psi(k_y,k_z)\nonumber \\
&=\sum_{-\sqrt{k_0^2-m^2}<k_z<\sqrt{\alpha^2-m^2}}\int dy\Psi_{k_z}^\dagger(y)\left(\begin{array}{cc}-i\partial_y & 0 \\ 0 & i\partial_y \end{array}\right)\Psi_{k_z}(y)\nonumber \\
&=\sum_{-\sqrt{k_0^2-m^2}<k_z<\sqrt{\alpha^2-m^2}}\int dy \Psi_{k_z}^\dagger(y)(-is^3\partial_y)\Psi_{k_z}(y) .
\end{align}
Then, we have the following effective action
\begin{align}
S&=\sum_{-\sqrt{k_0^2-m^2}<k_z<\sqrt{k_0^2-m^2}}\int d\tau\int dy\Psi_{k_z}^\dagger(y)\Big(\partial_\tau -is^3\partial_y\Big)\Psi_{k_z}(y)\nonumber \\
&=\sum_{-\sqrt{k_0^2-m^2}<k_z<\sqrt{k_0^2-m^2}}\int d\tau\int dy\bar{\Psi}_{k_z}(y)(\gamma^0\partial_\tau+i\gamma^1\partial_y)\Psi_{k_z}(y) ,
\end{align}
where $\gamma^0=s^1$, $\gamma^1=is^2$ and $\bar{\gamma}=\gamma^0\gamma^1=-s^3$. If we set $\gamma^2=-i\gamma^0$, we have
\begin{gather}
S=\sum_{-\sqrt{k_0^2-m^2}<k_z<\sqrt{k_0^2-m^2}}\int d^2x \bar{\Psi}_{k_z}(y)i\gamma^\mu \partial_\mu\Psi_{k_z}(y) ,
\end{gather}
where $\mu=1,2$, $\partial_0=\partial_2$ and $\{\gamma^\mu,\gamma^\nu\}=-\delta^{\mu\nu}$.

In order to find the anomaly, we couple both the charge gauge field and the spin gauge field to the above action
\begin{gather}
S=\sum_{-\sqrt{k_0^2-m^2}<k_z<\sqrt{k_0^2-m^2}}\int d^2x \bar{\Psi}_{k_z}(y)i\gamma^\mu(\partial_\mu+iA_\mu-iS_\mu \bar{\gamma})\Psi_{k_z}(y) .
\end{gather}
This action is invariant in the classical level under the following spin gauge transformation
\begin{gather}
\Psi_{k_z}\rightarrow e^{i\bar{\gamma}\theta}\Psi_{k_z},\; \bar{\Psi}_{k_z}\rightarrow \bar{\Psi}_{k_z}e^{i\bar{\gamma}\theta}, S_\mu\rightarrow S_\mu+\partial_\mu\theta.
\end{gather}

Now, the spin current is given by
\begin{gather}
Z=e^{-W[A,S]}=\int \mathcal{D}\bar{\Psi}\Psi e^{-S[\bar{\Psi},\Psi,A,S]}\\
j_{s}^\mu=\frac{\delta W[A,S]}{\delta S_\mu}=-\frac{1}{Z}\frac{\delta Z}{\delta S_\mu}=\sum_{-\sqrt{k_0^2-m^2}<k_z<\sqrt{k_0^2-m^2}}\langle \bar{\Psi}_{k_z}\gamma^\mu\bar{\gamma}\Psi_{k_z}\rangle .
\end{gather}
The calculation is completely the same as that of the valley Hall current, showing the anomaly and canceled by the anomaly inflow of the bulk.

\section{Generalization for the existence of two different regularization schemes in inversion symmetry-broken Weyl metals} \label{Appendix:general proof}

\begin{figure}
\begin{subfigure}[b]{0.2\textwidth}
\includegraphics[width=\textwidth]{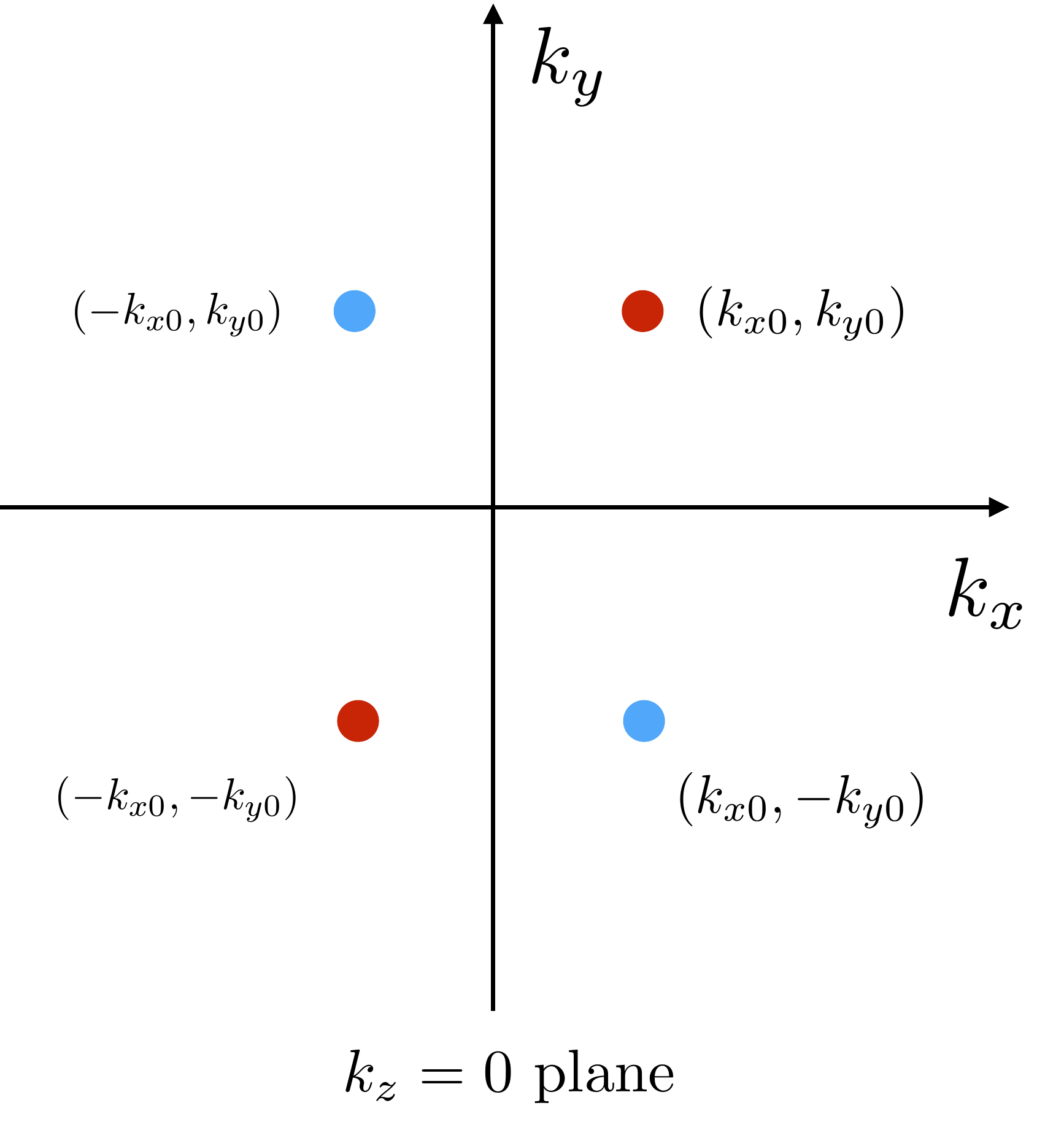}
\caption{Four Weyl points}\label{ValleyFermiArc}
\end{subfigure}
~
\begin{subfigure}[b]{0.2\textwidth}
\includegraphics[width=\textwidth]{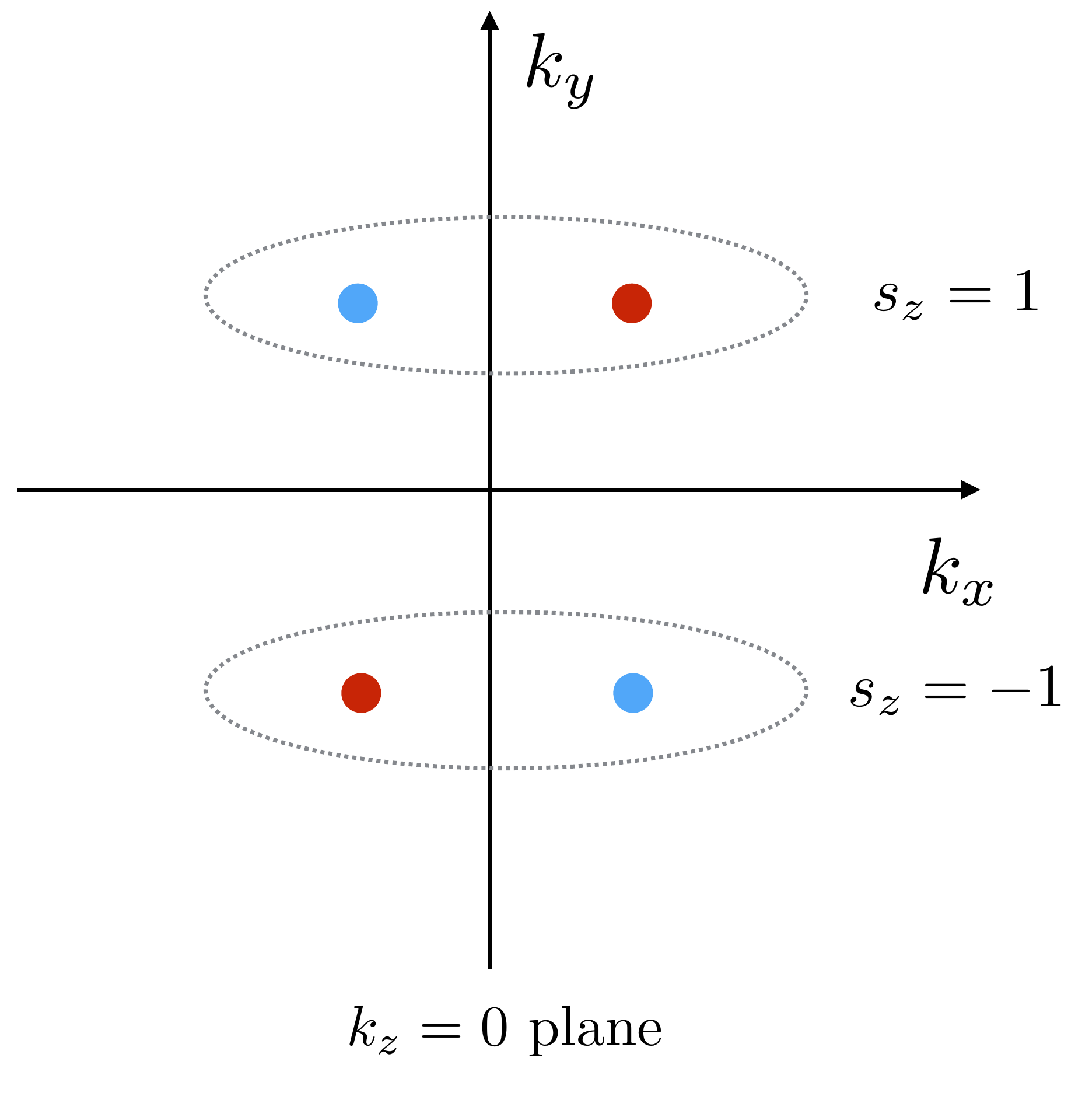}
\caption{Four Weyl points in the $s_z$ basis}\label{ValleyFermiArc}
\end{subfigure}
~
\begin{subfigure}[b]{0.2\textwidth}
\includegraphics[width=\textwidth]{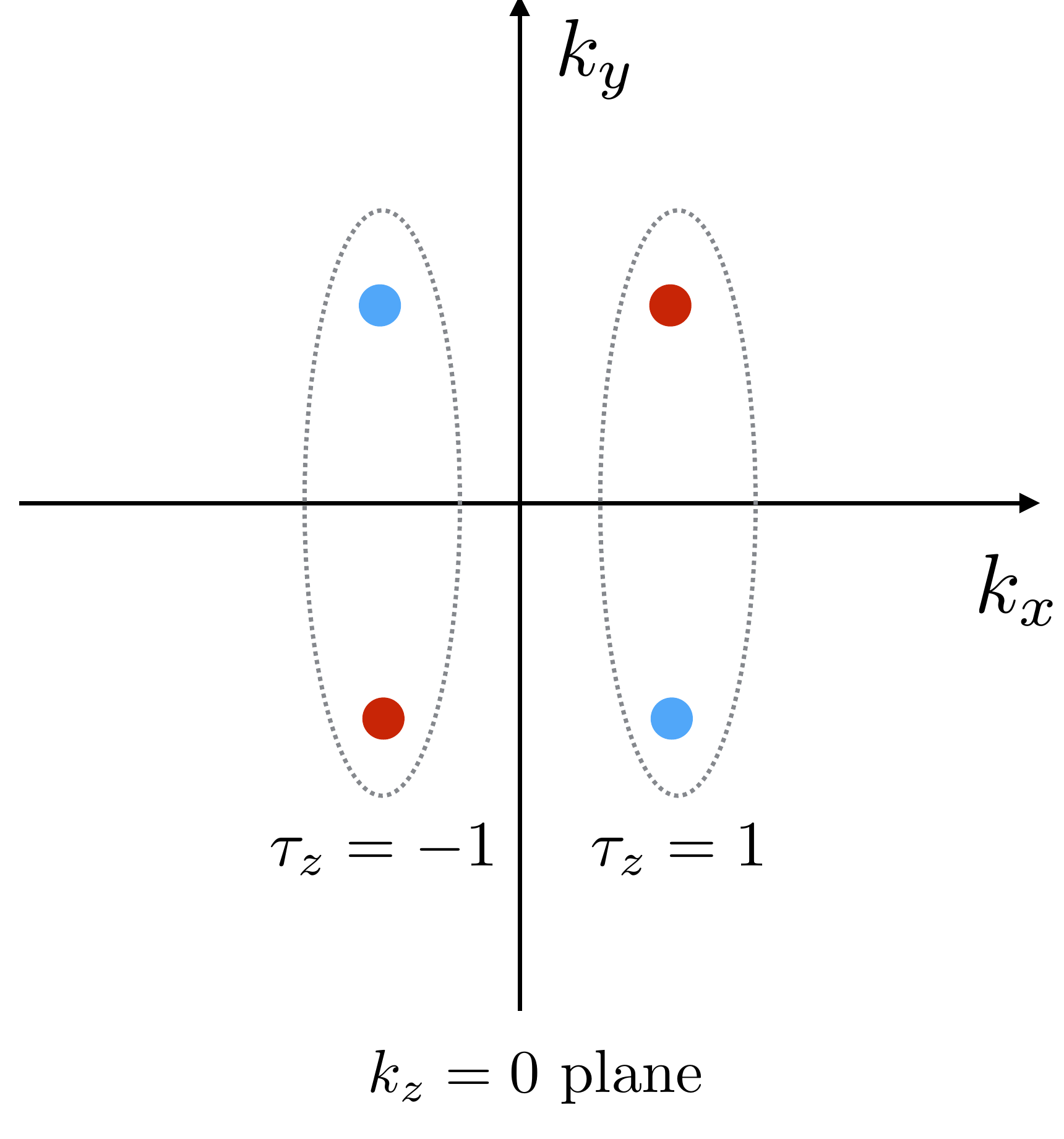}
\caption{Four Weyl points in the $\tau_z$ basis}\label{ValleyFermiArc}
\end{subfigure}
\caption{Four Weyl points}\label{4Weyl}
\end{figure}

Consider the Hamiltonian which have four Weyl points, shown in Fig. \ref{4Weyl} for example. In the diagonalized basis we obtain
\begin{align}
\mathcal{H}&=\sum_{k}\Psi^{\dagger}(k)diag(-E_{k_{x0},k_{y0}},E_{k_{x0},k_{y0}},-E_{-k_{x0},k_{y0}},E_{-k_{x0},k_{y0}},-E_{k_{x0},-k_{y0}},E_{k_{x0},-k_{y0}},-E_{-k_{x0},-k_{y0}},E_{-k_{x0},-k_{y0}})\Psi(k)\nonumber \\
&\equiv \sum_{k}\Psi^\dagger(k)H(k)\Psi(k) ,
\end{align}
where $E_{\pm k_{x0},\pm k_{y0}}=\sqrt{(k_{x}\mp k_{x0})^2+(k_y\mp k_{y0})^2+k_z^2}$. This diagonalized Hamiltonian commutes with two matrices; $\mathcal{O}_{s_z}$ and $\mathcal{O}_{\tau_z}$ given by
\begin{align}
\mathcal{O}_{s_z}=\left(\begin{array}{cccccccc} 1&0&0&0&0&0&0&0 \\ 0&1&0&0&0&0&0&0 \\ 0&0&1&0&0&0&0&0 \\ 0&0&0&1&0&0&0&0 \\ 0&0&0&0&-1&0&0&0 \\ 0&0&0&0&0&-1&0&0 \\ 0&0&0&0&0&0&-1&0\\0&0&0&0&0&0&0&-1\end{array}\right),\; \mathcal{O}_{\tau_z}=\left(\begin{array}{cccccccc} 1&0&0&0&0&0&0&0 \\ 0&1&0&0&0&0&0&0 \\ 0&0&-1&0&0&0&0&0 \\ 0&0&0&-1&0&0&0&0 \\ 0&0&0&0&1&0&0&0 \\ 0&0&0&0&0&1&0&0 \\ 0&0&0&0&0&0&-1&0\\0&0&0&0&0&0&0&-1\end{array}\right) .
\end{align}
Here, $\tau$ and $s$ do not mean the valley and spin necessarily but denote two different quantum numbers.

One can find that these different two observables turn out to commute with the Hamiltonian for any inversion symmetry-broken Weyl metals. If we classify the energy bands according to the eigenvalues of $\mathcal{O}_{\tau_z}$ and $\mathcal{O}_{s_z}$, we find
\begin{gather}
\Big\{\begin{array}{cc} s_z=1: &diag (-E_{k_{x0},k_{y0}},E_{k_{x0},k_{y0}},-E_{-k_{x0},k_{y0}},E_{-k_{x0},k_{y0}}) \\ s_z=-1: & diag (-E_{k_{x0},-k_{y0}},E_{k_{x0},-k_{y0}},-E_{-k_{x0},-k_{y0}},E_{-k_{x0},-k_{y0}}) \end{array} , \\
\Big\{\begin{array}{cc} \tau_z=1: & diag (-E_{k_{x0},k_{y0}},E_{k_{x0},k_{y0}},-E_{k_{x0},-k_{y0}},E_{k_{x0},-k_{y0}}) \\ \tau_z=-1: & diag (-E_{-k_{x0},k_{y0}},E_{-k_{x0},k_{y0}},-E_{-k_{x0},-k_{y0}},E_{-k_{x0},-k_{y0}}) . \end{array}
\end{gather}
Each sector is composed of two Weyl points.

For one $s_z$ sector, one can apply the unitary transformation to the Hamiltonian which gives the following transformed one
\begin{gather}
\tilde{H}(k)=\Big(\begin{array}{cc}\tilde{H}_{s_z=1}(k) & 0 \\ 0 & \tilde{H}_{s_z=-1}\end{array}\Big)=U_{s_z}(k)H(k)U^\dagger_{s_z}(k),\; U_{s_z}=\left(\begin{array}{cc}U_{s_z=1}|_{4\times 4} & 0 \\ 0 & U_{s_z=-1}|_{4\times 4} \end{array}\right)\\
\tilde{H}_{s_z}(k)=-i\gamma^0(\gamma^1k_x+\gamma^2(k_y-s_zk_{y0})+\gamma^3k_z-s_z\gamma^1\gamma^5k_{x0}) ,
\end{gather}
where $\gamma^0=\Big(\begin{array}{cc}0 & 1 \\ 1 & 0\end{array}\Big)$, $\gamma^i=i\Big(\begin{array}{cc}0 & -\sigma^i \\ \sigma^i & 0\end{array}\Big)$ with $i=1,2,3$ which satisfy $\{\gamma^{\mu},\gamma^\nu\}=2\delta^{\mu\nu}1_{4\times 4}$ and $\gamma^5=-\gamma_0\gamma_1\gamma_2\gamma_3=\Big(\begin{array}{cc}1 & 0 \\ 0& -1\end{array}\Big)$. Here, $\sigma^i$ represent Pauli matrices. The representation of gamma matrices is the Weyl one in the Euclidean signature. One can always find the unitary transformation $U_{s_z}$ which gives gamma matrices in this Weyl representation. Since we changed the basis by the unitary transformation, we need to change the representation of $\mathcal{O}_{s_z}$ and $\mathcal{O}_{\tau_z}$ coherently; $\tilde{\mathcal{O}}_{i}=U_{s_z}\mathcal{O}_{i}U_{s_z}^\dagger$. Recall that $\mathcal{O}_{s_z}$ commutes with $U_{s_z}$, and thus the representation of $\mathcal{O}_{s_z}$ is not changed; $\tilde{\mathcal{O}}_{s_z}=\mathcal{O}_{s_z}$. How about the representation of $\mathcal{O}_{\tau_z}$? Interestingly, $\tilde{\mathcal{O}}_{\tau_z}$ is also the same as $\mathcal{O}_{\tau_z}$; $\tilde{\mathcal{O}}_{\tau_z}=\mathcal{O}_{\tau_z}$. This can be verified easily. An important point is that $\tilde{\mathcal{O}}_{\tau_z}=\Big(\begin{array}{cc}\gamma^5 & 0 \\ 0 & \gamma^5\end{array}\Big)=\gamma^5\otimes 1_{\tau}$. As a result, the transformation $e^{i\tilde{\mathcal{O}}_{\tau_z}\theta}$ becomes anomalous in this representation (chiral anomaly).

There can exist other representations of gamma matrices even though the Hamiltonian $\tilde{H}(k)$ is the same. If we denote another representation of gamma matrices as $\bar{\gamma}^\mu$, it should satisfy the following properties:
\begin{gather}
\gamma^0\gamma^i=\bar{\gamma}^0\bar{\gamma}^i\;(i=1,2,3),\; \{\bar{\gamma}^\mu,\bar{\gamma}^\nu\}=2\delta^{\mu\nu}1_{4\times 4}\\
\bar{\gamma}^5=-\bar{\gamma}^0\bar{\gamma}^1\bar{\gamma}^2\bar{\gamma}^3 .
\end{gather}
From the properties of $\gamma^0\gamma^i=\bar{\gamma}^0\bar{\gamma}^i$ and $\{\bar{\gamma}^\mu,\bar{\gamma}^\nu\}=2\delta^{\mu\nu}1_{4\times 4}$, one can find the following identity; $\bar{\gamma}^5=\gamma^5$. Therefore, we obtain $\tilde{O}_{\tau_z}=\gamma^5\otimes 1_{\tau}=\bar{\gamma}^5\otimes 1_\tau$. As a result, any $\tilde{O}_{\tau_z} $ related transformation is still anomalous in any other representations. One may consider additional unitary transformations such as $\tilde{U}_{s_z}(k)$; $\tilde{\tilde{H}}(k)=\tilde{U}_{s_z}(k)\tilde{H}(k)\tilde{U}_{s_z}(k)$. Even in this case, $\tilde{\tilde{O}}_{\tau_z}$ is still anomalous since anti-unitary relation is preserved under the unitary transformation. Therefore, we conclude that if we choose gamma matrices which commute with $\mathcal{O}_{s_z}$, $\mathcal{O}_{\tau_z}$ should be proportional $\gamma^5$. This is also true for the opposite case: if we choose $\gamma$ matrices commuting with $\mathcal{O}_{\tau_z}$, then $\mathcal{O}_{s_z}$ is proportional to $\gamma^5$, which should be anomalous. In our specific model for inversion symmetry-broken Weyl metals, $\tau$ and $s$ correspond to valley and spin, respectively.

\end{document}